\documentclass[10pt]{article}
\usepackage{amsmath,amssymb}
\usepackage{amsthm}

\batchmode

\newtheorem{prop}{Proposition}

\newcommand{\pr}{\noindent{\bf Proof}. }
\newcommand{\rem}{\noindent{\bf Remark}. }
\newcommand{\res}{\noindent{\bf Remarks}. }

\newcommand{\pa}{\partial}

\newcommand{\const}{\textrm{const}}

\newcommand{\hs}{ \hspace{1cm}}
\newcommand{\tk}{\bbT^{-k}_{N -k}}
\newcommand{\tz}{\bbT^0_{N -k}}
\newcommand{\B}{\Big}
\newcommand{\be}{\begin{equation}}
\newcommand{\ee}{\end{equation}}
\newcommand{\ran}{ \mathrm{ ran }}

\newcommand{\sZ}{\mathsf{Z}}

\newcommand{\sx}{\mathsf{x}}
\newcommand{\sX}{\mathsf{X}}

\newcommand{\al}{\alpha}
\newcommand{\De}{\Delta}
\newcommand{\de}{\delta}

\newcommand{\Ga}{\Gamma}

\newcommand{\la}{\lambda}

\newcommand{\Si}{\Sigma}

\newcommand{\cA}{{\cal A}}

\newcommand{\cH}{{\cal H}}

\newcommand{\cG}{{\cal G}}
\newcommand{\cM}{{\cal M}}

\newcommand{\cQ}{{\cal Q}}

\newcommand{\cZ}{{\cal Z}}

\newcommand{\bbR}{{\mathbb{R}}}
\newcommand{\bbZ}{{\mathbb{Z}}}

\newcommand{\bbT}{{\mathbb{T}}}

\begin{document}

\title{Covariant  Axial Gauge}
\author{ 
J. Dimock
\thanks{dimock@buffalo.edu}\\
Dept. of Mathematics \\
SUNY at Buffalo \\
Buffalo, NY 14260 }
\maketitle

\begin{abstract}
We  consider    abelian gauge theories on a lattice   and develop properties  of  an axial gauge that is 
covariant   under lattice symmetries.     Particular  attention is paid  to a version that behaves nicely  under  block  averaging renormalization group transformations.   
\end{abstract}

 \section{Introduction}

  Gauge quantum field theory    can be formulated  in  various gauges.   Prominent choices are
  the axial gauge in which a   component  of the gauge field  is set to zero  and   covariant gauges  like Feynman  or
 Landau.   The axial gauge  is good for defining the theory and  exhibiting the positivity of the action.  The covariant gauges
 are good for  ultraviolet regularity and  exhibiting  the space-time symmetries of the theory.

 Balaban  in his studies of      renormalization group methods  for lattice gauge theories    found a way to exploit some   good 
 properties  of both types of gauges   \cite{Bal84a}, \cite{Bal84b},  \cite{Bal85b}.  In this  formulation the gauge field is a function 
 on  bonds in the lattice,   and the axial gauge was realized by setting  the field to zero on certain trees.   
 However    the axial gauge still  spoiled the space-time covariance.      When  he came to applying these methods  to  pure  Yang Mills in  $d=3,4$   \cite{Bal87}, \cite{Bal88a}  Balaban   found
that  this  was  a significant obstacle.    Instead  he developed a  covariant  axial  gauge in which he averaged over various trees to 
regain the covariance.     However  details   about  taking over the results  of   \cite{Bal84a}, \cite{Bal84b}, \cite{Bal85b}
were  absent.    Furthermore  in the Yang-Mills  papers  he used an exponential gauge fixing rather than the original  delta function gauge fixing.   
 
 In this paper  we   reconsider this covariant  axial  gauge  with  delta function gauge fixing,    and    establish    results     from   \cite{Bal84a}, \cite{Bal84b}, \cite{Bal85b}  for this case.     Our  purpose is to use them  in an analysis   of     ultraviolet problems  for   scalar  QED  in dimension $d=3$  \cite{Dim14}.    Scalar  QED was originally studied by  Balaban   \cite{Bal82a}, \cite{Bal82b},  \cite{Bal83a}, \cite{Bal83b}.     Some results were  extended to the abelian Higgs model   by    Balaban, Imbrie, and Jaffe  
 \cite{BIJ85},  \cite{BIJ88}.   See also  \cite{BaJa86},  \cite{Imb86}  for  further      discussion of these problems.

 In view of the intended application  we  mainly  work in    dimension  $d=3$,   but really   the results are not specific to any dimension.   
 In section  \ref{gauges}  we  develop the covariant axial gauge for the  free electromagnetic field  on unit lattice cube.
 In the remainder of the    paper,  section \ref{reform},   we   extend these results to a toroidal  lattice  with  arbitrarily small lattice spacing.      By  scaling this is equivalent  to   a unit lattice with a large volume.   The  results of section \ref{gauges}   
do not  give  good bounds in this case.    This is a case of a   massless model  in a large volume,   and this  is  just the  
arena for  
 renormalization group methods.   We  show  how   to  implement  the covariant axial gauge in a way compatible with block
 averaging renormalization group methods.

 \section{Axial gauges}   \label{gauges}

\subsection{gauge fixing on a tree}   \label{two-one}

Consider  an abelian gauge theory  on  a  finite unit   lattice  of dimension $d=2,3$;  specifically for an odd integer  $L$  on  the  square or     cube   
\be B(0) =   [- L/2,L/2]^d \cap    \bbZ^d   \ee
centered on the origin.       The gauge field  $A(b)  = A(x,x')$  is  an  $\bbR^d$ valued function on bonds  (=nearest neighbor pairs)  in  $B(0)$ which satisfies   $A(x',x) = - A(x,x')$.
The field strength  is defined on plaquettes  (= squares)   and is  
\be    dA( p)   =  \prod_{b \in \pa p}   A(b)  \ee
This is  invariant  under  gauge transformation  $A^{\la} = A- \pa \la$.   The action is  $\frac12 \| dA \|^2 =\frac12  \sum_p  | d A(p)|^2 $ and  we   are interested  in integrals of the form 
\be    \label{formal} 
 \int   f(A)   \exp \B( - \frac12 \| dA \|^2  \B)  DA        \hs       DA  =  \prod_{b  \in B(0)}  A(b)
 \ee
Here     $f( A)$ is assumed  bounded on compacts  and  gauge invariant, but with no particular decay at infinity. 

The integral is  not  convergent since  $dA$ has a large null space.   The axial gauge is the remedy.   We  first  explain the  tree axial gauge.   Let  
$\Ga_{0x}   $  be  the   rectilinear   path  in  the lattice  from  $0$ to $x$  obtained
 by successively increasing each coordinate to
its final value.   Thus  in $d=3$, $\Ga_{0x}$ is the path
\be    \label{gyx}
\Ga_{0x}  = \B[ (0,0,0),\     (x_1,0,0),\   (x_1,x_2, 0),\     (x_1,x_2, x_3)  \B]  
\ee
Introduce  new  variables    
$\tau^0  A$   defined on lattice sites  $ x \in B(0),   x  \neq  0 $   by 
\be
   (\tau^0 A) (x)   =  A( \Ga_{0x}  )       \hs     A( \Ga )   =  \sum_{b \in \Ga }  A(b)
\ee
Note  that  under  a  gauge  transformation   we  have  
\be 
 (\tau^0 A ^{\la}) (x)  =    (\tau^0 A) (x)   -     \sum_{b \in \Ga_{0x} }   \pa \la (b)  
 =    (\tau^0 A) (x)   - (  \la(x)  - \la(0) )   
\ee

\begin{figure}[ht]  
\centering
\begin{picture}(100,100)(0,0)
  \put(0,0){\circle*{3}}
 \put(20,0){\circle*{3}}
 \put(40,0){\circle*{3}}
 \put(60,0){\circle*{3}}
  \put(80,0){\circle*{3}}
\ 
  \put(0,20){\circle*{3}}
 \put(20,20){\circle*{3}}
 \put(40,20){\circle*{3}}
 \put(60,20){\circle*{3}}
  \put(80,20){\circle*{3}}
\
  \put(0,40){\circle*{3}}
 \put(20,40){\circle*{3}}
 \put(40,40){\circle*{3}}
 \put(60,40){\circle*{3}}
  \put(80,40){\circle*{3}}
\
  \put(0,60){\circle*{3}}
 \put(20,60){\circle*{3}}
 \put(40,60){\circle*{3}}
 \put(60,60){\circle*{3}}
  \put(80,60){\circle*{3}}
\
  \put(0,80){\circle*{3}}
 \put(20,80){\circle*{3}}
 \put(40,80){\circle*{3}}
 \put(60,80){\circle*{3}}
  \put(80,80){\circle*{3}}
\
\thinlines
\put(0,40){\line(1,0){80}}
\put(0,0){\line(0,1){80}}
\put(20,0){\line(0,1){80}}
\put(40,0){\line(0,1){80}}
\put(60,0){\line(0,1){80}}
\put(80,0){\line(0,1){80}}
\
\put(80,40){\vector(1,0){15}}
\put(100,38){\text{$x_1$}}
\put(40,80){\vector(0,01){15}}
\put(40,100){\text{$x_2$}}
\put(32,32){\text{0}}
\end{picture}
\caption{The  tree $T$ for  $d=2, L=5$   }
\label{bounty1}
\end{figure}
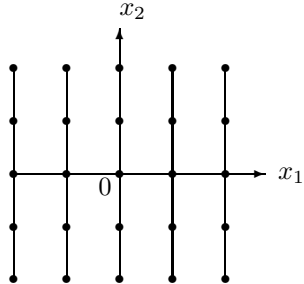

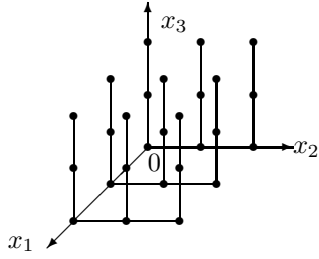
\begin{figure}[ht]  
\centering
\begin{picture}(100,100)(0,0) 
 \put(0,0){\circle*{3}}
\put(14,14){\circle*{3}}
\put(28,28){\circle*{3}}
\put(20,0){\circle*{3}}
\put(34,14){\circle*{3}}
\put(48,28){\circle*{3}}
\put(40,0){\circle*{3}}
\put(54,14){\circle*{3}}
\put(68,28){\circle*{3}}
\ 
\put(0,20){\circle*{3}}
\put(14,34){\circle*{3}}
\put(28,48){\circle*{3}}
\put(20,20){\circle*{3}}
\put(34,34){\circle*{3}}
\put(48,48){\circle*{3}}
\put(40,20){\circle*{3}}
\put(54,34){\circle*{3}}
\put(68,48){\circle*{3}}
\
\ 
\put(0,40){\circle*{3}}
\put(14,54){\circle*{3}}
\put(28,68){\circle*{3}}
\put(20,40){\circle*{3}}
\put(34,54){\circle*{3}}
\put(48,68){\circle*{3}}
\put(40,40){\circle*{3}}
\put(54,54){\circle*{3}}
\put(68,68){\circle*{3}}
\thinlines
\put(0,0){\line(1,1){28}}
\put(0,0){\line(1,0){40}}
\put(14,14){\line(1,0){40}}
\put(28,28){\line(1,0){40}}
\
\put(0,0){\line(0,1){40}}
\put(20,0){\line(0,1){40}}
\put(40,0){\line(0,1){40}}
\put(14,14){\line(0,1){40}}
\put(34,14){\line(0,1){40}}
\put(54,14){\line(0,1){40}}
\put(28,28){\line(0,1){40}}
\put(48,28){\line(0,1){40}}
\put(68,28){\line(0,1){40}}
\
\put(0,0){\vector(-1,-1){10}}
\put(-25,-10){\text{$x_1$}}
\put(68,28){\vector(1,0){15}}
\put(83,26){\text{$x_2$}}
\put(28,68){\vector(0,1){15}}
\put(33,73){\text{$x_3$}}
\put(28,19){\text{0}}
\end{picture}
\caption{Part of the  tree $T$ for  $d=3, L=5$   }
\label{bounty2}
\end{figure}

Let   $T$  be the  oriented tree  consisting of all bonds that occur in  any    $(\tau^0 A)(x)$.       See  figures   \ref{bounty1}  and    \ref{bounty2}.    
The  tree axial gauge   means that in the formal  integral  (\ref{formal})     we  set   $A(b) =0$ for all  $b \in T$. 
If  $<x,x'>$ is on the tree then  $A(x,x')   = ( \tau^0 A)(x') -   ( \tau^0 A)(x)$,   so it is equivalent to set  $(\tau^0 A)(x)=0$
for all $x$. 

This mutilation  can be motivated by a  
 Fadeev- Popov argument.   Let   $Q \la$ be the average
 \be  
 Q \la   =  L^{-d}   \sum_{x \in B(0) }  \la(x) 
 \ee
 and define        
 \be   \de (   \tau^0 A  )  =  \prod_{x  \neq   0}  \de \B(  (\tau^0 A )(x)     \B)   \hs    D \la =   \prod_{x   \in B(0)}   d (\la(x) )
\ee
 Start with the identity 
 \footnote{For a general  theory of integrals of the form  $\int \de( \phi(x)) f(x)  dx  $ see Gelfand and Shilov  \cite{GeSh64}.   For us   $\phi$ will  always be linear or affine.}  
\be        \int      \de (Q \la  )    \de (  \tau^0 A ^{\la}  )\  D \la      =     \int    \de (Q \la  )   \de \B(   \tau^0 A  -   (\la-   \la(0)  \B)\  
  D \la    =   \const     
\ee
This can be seen by making the change of variables   $\{  \la(x) \} $   to   $\{ \la(x) - \la(0)   \}_{x \neq  0},  Q \la $ which has
a constant Jabobian.   
Insert this under the integral sign  in     (\ref{formal})  and change the order of integration to obtain up to a constant multiple
\be      \label{formal2} 
\int   \left[ \int   f(A)    \de (  \tau^0 A ^{\la}  ) 
 \exp \B( - \frac12 \| dA \|^2  \B)   \  DA  \right]  \     \de (Q \la  )    D \la 
 \ee
Now  in  the  bracketed expression     make the change of variables  $A  \to  A^{- \la}$.    Since  $f(A)$ and  $d A$ are gauge
invariant  we  get the same  thing  with   $\la=0$.   Take the bracketed expression  outside the $\la$ integral  and then   throw   away the  remaining  infinite $\la$
integral.  We   end with the  desired   expression   
\be      \label{start1}       \int  
  f(A)      \     \de (  \tau^0 A   )  \exp \B( - \frac12 \| dA \|^2  \B)\   DA  \  
 \ee

\subsection{covariant axial gauge} 
For the covariant  axial gauge we  average over the ordering of the coordinates   in  the path  from $0$ to $x$.  
Let  $\pi$  be a permutation  of  $(1,2,  \dots, d)$ and  let   
   $\Ga^{\pi}_{0x}$  be all  rectilinear  paths  from  $0$ to $x$ in  which the coordinates   are 
taken to their final values    in the order determined by  $\pi$.  
 If $\pi$ is the identity  then  $\Ga^{\pi}_{0x} =\Ga_{0x}$    
We     replace   $\tau^0$   by  an average over   permutations  
\be 
( \tau A) (x)  = \frac{1}{  d!  }     \sum_{  \pi }   A  ( \Ga^{\pi}_{0x} )  =  \frac{1}{|G(0,x)|}  \sum_{\Ga \in G(0,x)}  A(\Ga) 
\ee
    In the second form  we   let $G(0,x)$ stand for the set of
all   $\Ga^{\pi}_{0x} $   and   $|G(0,x)|$ is again $d!$.

This is covariant in the following  sense.   Let $r$  be a     lattice symmetry leaving the origin fixed  and  let  $A_r(b)  = A(r^{-1} b)$. Then     
  $A_r(\Ga)  = A(r^{-1} \Ga) $  and   $r G(0,x)  =  G(0,rx)$  imply that  
  \be
  ( \tau  A_r )(x)  =    ( \tau A)_r(x) \equiv   ( \tau A)(r^{-1}x)
\ee
It follows that
\be    \de ( \tau A_r)  =  \de ( \tau A)
\ee
We  still have   
  \be   
 (\tau A ^{\la}) (x) 
 =    (\tau A) (x)   -  ( \la(x)  - \la(0) )    
\ee
Hence   by exactly the same  formal     argument that led  from  (\ref{formal})  to        (\ref{start1})  we can go instead from  (\ref{formal}) to  
\be        \label{start2}       \int      \de (  \tau  A   ) 
  f(A)   \exp \B( - \frac12 \| dA \|^2  \B)  \ DA  \ 
 \ee
This is our starting point. 

The next  result shows that  the gauge fixing has done its  job  and   the 
the integrals   (\ref{start1}) and (\ref{start2})  are finite.

 \begin{prop}   {  \  }  \label{santora}  On the cube  $B(0)$
\begin{enumerate} 
\item  
If     $\tau^0 A  =  0   $   and   $dA  = 0$     then  $A = 0$.
\item   If     $\tau A  =  0   $   and   $dA  = 0$     then  $A = 0$.
\item    There      exists a constant  $C$ (depending on L) such that   if   either       $\tau^0 A  =  0   $  or       $\tau A  =  0   $  
\be      \| dA \|^2   \geq  C   \| A  \|^2  \ee
\item   If  $f(A)$ is exponentially bounded the integrals   (\ref{start1}),(\ref{start2}) exist. 
\end{enumerate}
\end{prop}
\bigskip

\pr    For the first   we  use the principle that if  $dA(p) =0$   and  we know that   $A(b) =0$ for 
three of the  bonds in $\pa p$ then  $A(b)=0$ for the fourth bond.    Hence starting at the origin and
working outward we deduce    that  $A(b) =0$  for   bonds  $b$  not on the tree $T$,  and hence for  all $b$.
For the second  point   note    that   
\be  \tau A - \tau^0 A  =     \frac{1}{ d! }     \sum_{\pi  }   A  ( \Ga^{\pi}_{0x}  )  -  A (\Ga_{0x})
=  0
\ee
since   $ \Ga^{\pi}_{0x}   - \Ga_{0x} $ is  a closed path  and  $dA =0$  means   the integral  of  $A$   over closed paths vanishes. 
Hence   $\tau A =0$ implies  $\tau^0 A =0$   and again  $A=0$.
 The third follows since a  positive definite quadratic form on a finite dimensional  vector space is bounded  below on
 the unit sphere.  The fourth follows from the third.

\subsection{parametrization}  \label{this}

We   can    carry out integrals  in the  axial gauge  by  introducing new coordinates which include $\tau A$.
We  have        
\be    
\begin{split}
\int \de ( \tau A)  f(A) DA   =  &  \int_{\ker \tau  \times  (\ker \tau)^{\perp} }  \de (  \tau A_2 ) f(A_1 + A_2 )\   DA_1 DA_2    \\ 
=   &  \det\B( \tau\  | ( \ker \tau )^{\perp} \B) ^{-1}   \int_{\ker \tau  \times   \bbR^{B(0) - \{ 0 \} }  }  \de ( T ) f(A_1 + \tau^{-1}T )\   DA_1 DT \\
= &   \det\B( \tau\  | ( \ker \tau )^{\perp} \B) ^{-1}   \int_{ \ker \tau}  f(A_1 )\   DA_1\\
\end{split}
\ee
We have made a change of variables    $T  = \tau A_2  $  using that     $\tau :     (\ker \tau)^{\perp}    \to  \bbR^{B(0) - \{ 0 \} }  $  is a bijection.
Indeed it     is  injective since the kernel is zero and    it  is     surjective  since   $(\tau (d \la))(x)  = \la(x) - \la(0)$.  
Finally the integral   $ \int_{ \ker \tau}  f(A_1 )\   DA_1$
is evaluated by picking any orthonormal basis for   $\ker  \tau$   and reducing it to an integral over  $\bbR^n$ where 
$n = \dim ( \ker \tau )$.

 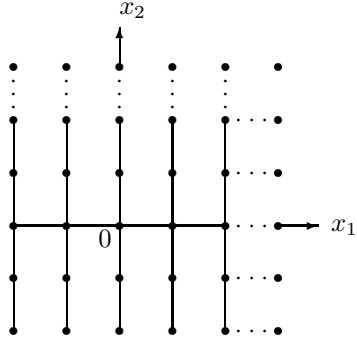
\begin{figure}[ht]  
\centering
\begin{picture}(100,100)(0,0)
  \put(0,0){\circle*{3}}
 \put(20,0){\circle*{3}}
 \put(40,0){\circle*{3}}
 \put(60,0){\circle*{3}}
  \put(80,0){\circle*{3}}
    \put(100,0){\circle*{3}}

\ 
  \put(0,20){\circle*{3}}
 \put(20,20){\circle*{3}}
 \put(40,20){\circle*{3}}
 \put(60,20){\circle*{3}}
  \put(80,20){\circle*{3}}
   \put(100,20){\circle*{3}}

\
  \put(0,40){\circle*{3}}
 \put(20,40){\circle*{3}}
 \put(40,40){\circle*{3}}
 \put(60,40){\circle*{3}}
  \put(80,40){\circle*{3}}
 \put(100,40){\circle*{3}}

\
  \put(0,60){\circle*{3}}
 \put(20,60){\circle*{3}}
 \put(40,60){\circle*{3}}
 \put(60,60){\circle*{3}}
  \put(80,60){\circle*{3}}
 \put(100,60){\circle*{3}}

\
  \put(0,80){\circle*{3}}
 \put(20,80){\circle*{3}}
 \put(40,80){\circle*{3}}
 \put(60,80){\circle*{3}}
  \put(80,80){\circle*{3}}
 \put(100,80){\circle*{3}}
 \  
\put(0,100){\circle*{3}}
\put(20,100){\circle*{3}}
\put(40,100){\circle*{3}}
\put(60,100){\circle*{3}}
\put(80,100){\circle*{3}}
\put(100,100){\circle*{3}}
\
\thinlines
\put(0,40){\line(1,0){80}}
\put(0,0){\line(0,1){80}}
\put(20,0){\line(0,1){80}}
\put(40,0){\line(0,1){80}}
\put(60,0){\line(0,1){80}}
\put(80,0){\line(0,1){80}}
\
\put(0,85){\circle*{1}}
\put(0,90){\circle*{1}}
\put(0,95){\circle*{1}}
\put(20,85){\circle*{1}}
\put(20,90){\circle*{1}}
\put(20,95){\circle*{1}}
\put(40,85){\circle*{1}}
\put(40,90){\circle*{1}}
\put(40,95){\circle*{1}}
\put(60,85){\circle*{1}}
\put(60,90){\circle*{1}}
\put(60,95){\circle*{1}}
\put(80,85){\circle*{1}}
\put(80,90){\circle*{1}}
\put(80,95){\circle*{1}}
\
\put(85,0){\circle*{1}}
\put(90,0){\circle*{1}}
\put(95,0){\circle*{1}}
\put(85,20){\circle*{1}}
\put(90,20){\circle*{1}}
\put(95,20){\circle*{1}}
\put(85,40){\circle*{1}}
\put(90,40){\circle*{1}}
\put(95,40){\circle*{1}}
\put(85,60){\circle*{1}}
\put(90,60){\circle*{1}}
\put(95,60){\circle*{1}}
\put(85,80){\circle*{1}}
\put(90,80){\circle*{1}}
\put(95,80){\circle*{1}}
\
\put(100,40){\vector(1,0){15}}
\put(120,38){\text{$x_1$}}
\put(40,100){\vector(0,01){15}}
\put(40,120){\text{$x_2$}}
\put(32,32){\text{0}}
\end{picture}
\caption{The  tree $T$ in the torus.    Points on opposite sides are identified.   }
\label{bounty3}
\end{figure}

      \subsection{torus}   \label{torus}

We now discuss how these developments can be extended to a torus.     
For the torus  we  again take the cube  $B(0)$,  but  now  include bonds joining points on opposite sides,  the dotted lines in   figure \ref{bounty3} .   We  cannot extend 
the tree to include these  bonds since that  would mean closing a  loop  which cannot be justified.    Working  with 
the old tree  we  gauge fix as before  and obtain   again 
\be        \label{start3}       \int      \de (  \tau^0  A   ) 
  f(A)   \exp \B( - \frac12 \| dA \|^2  \B)  \ DA  \ 
 \ee
However this  integral is still not convergent since there is no decay in   the gauge field on the new  bonds.

To fix this define a  vector   $\cQ^{\bullet}  A$  by   
\be   \label{bullet}
  (\cQ^{\bullet}  A)_{\mu}  =   L^{-d}  \sum_{x \in B(0)}   A( \Ga_{x, \mu} )   \ee
where  $\Ga_{x, \mu}$ is  the  path  around the torus through $x$ in the direction $e_{\mu}$.
We   insert a delta function  enforcing that   $ (\cQ^{\bullet}  A)_{\mu}  =0$   under the integral and 
obtain     a new starting point
\be        \label{start4}       \int   \de (  \cQ^{\bullet}  A)      \de (  \tau^0  A   ) 
  f(A)   \exp \B( - \frac12 \| dA \|^2  \B)  \ DA  \ 
 \ee
This  is not  gauge fixing.   Rather it is suppressing the contribution of torons  (Wilson lines),  something which presumably is inconsequential in the
infinite volume limit.

The integral is now convergent.   To see it we show that   $ \cQ^{\bullet}  A=0$ and  $\tau^0 A =0$  and   $dA =0$
imply that  $A =0$.  As before    $\tau^0 A =0$  and   $dA =0$ imply that   $A=0$  on all the bonds joining points  in  $T$.
For a new bond $<x, x+e_{\mu}>$,  the dotted lines in figure \ref{bounty3},   $dA=0$  and  $A=0$ on bonds joining points in $T$ imply that  
$A_{\mu} (x,x+ e_{\mu})  = c_{\mu}$ a constant.  Since  now     $ (\cQ^{\bullet}  A)_{\mu}  = c_{\mu}$  the constant must be  zero
and hence the result.

The reason  for working on a torus  is to increase  group of lattice symmetries.    We  have made  several special choices here
to spoil those symmetries.   This could be fixed by averaging over the various choices.   But this is not necessary  for our purposes.      In the renormalization group  approach  one works on a large torus   which is 
broken up into cubes.   On each  cube  we can use the covariant   gauge fixing  of section  \ref{two-one} to preserve
the symmetries of the effective interaction.   Only  in   the last
step  when  the  volume has shrunk to  a single cube  do we  need the torus version of gauge   fixing.  Here covariance  does  not matter  since after the final integral all the fields are gone.
   
  We proceed to  explain the renormalization group program in  more detail.

\section{Renormalization group}   \label{reform}

\subsection{orientation}

For  an abelian  gauge  theory in dimension $d=3$  we  take  a  fixed  large   $L$  and    introduce   toroidal lattices  
\be   \bbT^{-N}_M    =  L^{-N}   \bbZ^3/  L^M  \bbZ^3  
\ee
with spacing  $L^{-N}$ and volume  $L^{3M}$.

For  ultraviolet problems   we      start on  $\bbT^{-N}_0$  with  spacing  $L^{-N}$ and  unit volume  and   
 consider    formal    integrals like
\be     \label{munch1}   
 \int   f(\cA)   \exp \B( - \frac12 \| d\cA \|^2  \B)  D\cA        \hs       D\cA  =  \prod_{b  \in   \bbT^{-N}_0} d(  \cA(b)  )
 \ee
 The function $f(\cA)$ carries   the contribution of any   other fields  and is assumed gauge invariant.

The  general problem is  first  to make sense of this  integral   and   second    
 take the limit  $N \to \infty$ or  at least get bounds uniform in $N$.
Again the  solution to  the first  problem    is gauge fixing.   But if we  gauge fix on  a giant  tree and use it directly  we  do not
get  bounds uniform in $N$.    Instead we  gauge  fix  on  a hierarchical tree  which we  now explain.   This is more
compatible   with  renormalization group transformations which provide the solution to the  second  problem.  

First some  definitions.  Let   $\cA$  be a 
function on bonds on a lattice    $T^{-k}_*$ with spacing  $L^{-k}$ and arbitrary volume.
      We  define  an  averaged    field   $\cQ A$  defined    on  oriented    bonds
 $<y,  y + L^{-k+1} e_{\mu}> $  in      $\bbT^{-k+1}_* $ by     (for reverse  oriented bonds take minus this)  
\be   
(\cQ \cA) (y,  y + L^{-k+1} e_{\mu} )  
=  \sum_{x \in B(y)  } L^{-4}    \cA(  \Ga_{x,  x +  L^{-k+1}  e_{\mu} } )   
\ee
Here  $B(y)$  is  a block with  $L$ sites in each direction centered on $y$,   and    $ A(\Ga)  = \sum_{b \in \Ga}  A(b)$ is an unweighted sum over  bonds $b$ in  $T^{-k}_*$.  This  is  scale invariant.   We also   define  the $n$ fold composition  $\cQ_n  =   \cQ  \circ  \cdots  \circ  \cQ$ 
which takes fields on  $\bbT^{-k}_*$ to fields on $\bbT^{-k+n}_*$  and is given by 
\be   
(\cQ_n \cA) (y,  y + L^{-k+n} e_{\mu} )  
=  \sum_{x \in B^n(y)  } L^{-4n}    \cA(  \Ga_{x,  x +  L^{-k+n}  e_{\mu} } )     
\ee
Here  $B^n(y)$  is  a block with  $L^n$ sites in each direction centered on $y$

Also on   any  lattice     $T^{-k}_*$  for  $x \in B(y)$  define        
\be  ( \tau \cA  ) (y,x)   =    \frac{1}{ d!   }     \sum_{\pi }   \cA  ( \Ga^{\pi}_{yx} )  =   \frac{1}{|G(y,x)|}  \sum_{\Ga \in G(y,x)} \cA(\Ga)  
\ee
Here  $ \Ga^{\pi}_{yx} $  is     the rectilinear  paths  from  $y$ to $x  \in B(y)  $ in  which the coordinates   are 
taken from  $y$   to their final values   $x$  in the order determined   by  $\pi$,   and  $G(y,x)$ is all such paths.
Note that  since   $\cA (\Ga)$  defined with  an unweighted sum    
we have   $(\pa \la)(\Ga_{yx} )  = L^k (\la(x) - \la(y))$  and hence  $ ( \tau  \pa \la   ) (y,x)  =    L^k (\la(x) - \la(y))$.
Note also that  
if     $r$ is a lattice symmetry then   $r G(y,x)  = G(ry,rx)$   and so   $ ( \tau  \cA_r  ) (y,x)  =   ( \tau \cA  ) (r^{-1}y,r^{-1}x) $.

Returning to our problem on  $\bbT_0^{-N}$   we  first  define a gauge fixing function 
 for  $\cA$   on  bonds in     $\bbT_0^{-N+j}$  for  $j=0, \dots,  N-1$   by             
\be   \label{heathrow} 
\de  (\tau \cA)    =  \prod_{y   } 
 \prod_{       x  \in B(y),  x \neq  y    } 
 \de \B((\tau \cA)(y,x) \B)   
\ee
where   $y  \in   \bbT^{-N+j+1}_0  $ and $x  \in \bbT^{-N+j}_0$.
This      satisfies    $\de  (\tau \cA_r)   =  \de  (\tau \cA)$.       
Then  for     a function   $\cA$   on bonds  in   $\bbT^{-N}_0$,  $\cQ_j \cA$ is a function on bonds in   $\bbT^{-N+j}_0$,
and we  define  the    gauge fixing  function 
 \be
   \de^{\sX}_N (\cA)   =      \prod_{j=0}^{N-1}    \de  (\tau  \cQ_j \cA )   
 \ee
 This   is also invariant under lattice symmetries.

 We  would like to    insert   $  \de^{\sX}_N   (\cA)  $ in the integral   (\ref{munch1}).   To motivate this we need:

 \begin{prop}  \label{three}  For $\cA$ on  $\bbT^{-N}_0$ the integral    
 \be
        \int    \de ( Q_N \la  )    \de^{\sX}_N (\cA^{\la})    D\la     \hs   D \la  = \prod_{x \in   \bbT^{-N}_0}  d( \la(x)) 
 \ee  is constant.  
\end{prop}   
\bigskip

\rem  In  general  $Q_k =  Q \circ \cdots   \circ  Q$ ($k$ times)   
 averages scalars   over blocks with  $L^{3k}$ sites in each direction, and is given by
\be 
  (Q_k f  )(y)   =  L^{-3k}  \sum_{x \in B^k(y)  }  f(x) 
     \ee
  So for  $\la$   on  $\bbT_0^{-N}$ we have  that    $Q_N \la $ is a single number equal to  the average of  $\la$ over the whole lattice .
\bigskip

\pr    We have    
\be   \label{sunset3}
\begin{split}
&   \int    \de ( Q_N \la  )   \de^{\sX}( \cA^{\la})  \      D  \la   
=     \int  \de (Q_N \la )   \prod_{j=0}^{N-1}   \de (\tau  \cQ_{j} \cA^{\la} )    D \la    \\
= &     \int  \de (Q_N \la ) \prod_{j=0}^{N-1}    \prod_{y_{j},x_{j}  }   \de  \B( (\tau \cQ_{j} \cA)(y_j, x_j)
 - L^{N-j}(\  \cQ_{j} \la(x_{j})   - \cQ_{j} \la(y_{j})\ ) \B)    \     D \la    \\
 \end{split}
\ee
The  interior product is over          
\be     y_j  \in    \bbT^{-N+j+1}_0   \hs     x_j   \in    \bbT^{-N+j}_0  \hs    x_j \in B(y_j)   \hs        x_j   \neq   y_j  \ee
   and  
we  have used the identities    $  \cQ_j  \pa \la   =    \pa Q_j \la  $
and        $(\tau  \pa Q_j\la  ) (y_j,x_j)   =   L^{N-j} (  Q_j \la (x_{j})   - Q_j\la(y_j))$.
  
  We  change variables   from  $\{\la(x) \}$   for   $x \in   \bbT^{-N}_0  $  to  
  \be   
  Q_N\la \ \   \textrm{  and  } \ \    \{ Q_j \la (x_{j})   - Q_j\la(y_j)  \}_{x_j \neq  y_j}      \ \ \ \   j=0, \dots, N-1    
 \ee 
      We  claim   
  this linear transformation is non-singular.    The number of variables  is the  same, namely  
  \be  (  L^{3N} - L^{(3N-1) }  ) +     \cdots  +   (L^{6}-L^3)  + (L^3-1)  + 1 = L^{3N} 
\ee
  so  it suffices to show  kernel  is  zero.  But  $Q_N \la   = Q  Q_{N-1} \la = 0$ 
  and   $ Q_{N-1} \la(x_{N-1})  -    Q_{N-1} \la(y _{N-1})      =0  $ for   $x_{N-1} \neq  y_{N-1}   $  imply   $ Q_{N-1} \la(x_{N-1})    =0  $ for all   $x_{N-1}$.  This is the same as  $(Q  Q_{N-2} \la)(y_{N-2}) =0 $ for all $y_{N-2} $.    Combine this with
  $ Q_{N-2} \la(x_{N-2})  -    Q_{N-2} \la(y_{N-2})      =0  $  for all  $x_{N-2} \neq  y_{N-2}$  and conclude that      $ Q_{N-2} \la(x_{N-2} )   =0  $ for all   $x_{N-2}$.
  Continue the argument until  we   get  to  $Q\la(y_0) =0$    and  $ \la(x_0)- \la(y_0)    =0$  for  $x_0 \neq  y_0$   to conclude  $\la(x_0) =0$
    for all $x_0$.

  Thus we  can write   in  (\ref{sunset3})  
  \be 
    D \la   =  \const  \
       d ( Q_N \la  )    \prod_{j=0}^{N-1}   \prod_{y_j,x_j }  \ d\B( (Q_j \la)(y_j)  -(Q_j \la)(x_j) \B) 
  \ee
 Carrying out    the integrals  in  (\ref{sunset3})    with     these variables yields a constant. 
     This completes the proof.   
\bigskip

 Using this result we again make a  Fadeev-Popov argument.    Insert  the integral 
   $   \int    \de ( Q_N \la  )    \de^{\sX}_N (\cA^{\la})    D\la  $  in  (\ref{munch1})   and   the    change the order of integration
   to  get    
   \be     \label{munch2}   
 \int  \left[     \int   f(\cA)    \de^{\sX}_N (\cA^{\la})   \exp \B( - \frac12 \| d\cA \|^2  \B)  D\cA    \right]    \de ( Q_N \la  )      D\la   
\ee
  Now   
  gauge away the $\la$  dependence  in the integral over $\cA$ and   then    throw away the infinite integral  over  $\la$ and any other constants.   
 This yields the gauge fixed integral     
\be     \label{formal4} 
 \int     f(\cA)   \de^{\sX}_N  (\cA)     \exp \B( - \frac12 \| d\cA \|^2  \B)  D \cA       
 \ee
 
 This is still not well-defined  because  of the toron problem  mentioned  in   section  \ref{torus}.  The remedy is the same.
 Define  
 \be    \cQ_N^{\bullet}  = \cQ^{\bullet}   \circ   \cQ_{N-1}  
 \ee
 Here  $\cQ^{\bullet}$ is defined as  in  (\ref{bullet})  but now on an $L^{-1}$ lattice instead of a unit lattice.    
 Insert    $\de  (    \cQ_N^{\bullet}   \cA  )$ under the integral sign to obtain  
 \be     \label{formal5} 
 \int  f(\cA)   \de  (    \cQ_N^{\bullet}   \cA  )   \de^{\sX}_N  (\cA)     \exp \B( - \frac12 \| d\cA \|^2  \B)  D \cA       
 \ee
This is our new starting point.   We  will see   in the next section   that it is well-defined.

\subsection{renormalization group   transformations}

We  next  explain  the renormalization group transformations.  For this  
it is convenient to work with unit lattice variables so we   start by  
scaling        up to the torus    $\bbT^0_N$ with
unit spacing and volume  $L^{3N}$.    If    $A$ is  field on $\bbT^0_N$  then
  \be  
\cA(b)    = A_{L^{-N}}(b)  \equiv  L^{ \frac{ N }{2}    } A(L^{N}b)   
 \ee
 is a field on  $\bbT^{-N}_0$.  We    substitute it into   the  original density  $f(\cA) \exp \left( - \frac12 \| d\cA \|^2  \right)    $ ,  use the fact  that  $\| d\cA \|^2$ is 
 invariant, and    get   a new  unit lattice density 
 \be 
   \rho_0 (   A )   =      F_0(A)    \exp \B( - \frac12 \| dA \|^2  \B)   \hs     F_0(A)   \equiv     f_{L^N}(  A ) \equiv    f(  A_{ L^{-N }}   ) 
 \ee

Starting with  $ \rho_0 $ on $\bbT^0_N$    we  generate
  the  integral   (\ref{formal5})   in a series  of steps.
We  successively define  densities  $  \rho_k$  on fields  in $ \bbT^0_{N-k}$   by block averaging.  
Given   $\rho_k$   we first   define    $\tilde  \rho_{k+1} $  on   fields  $A_{k+1}$ on     $ \bbT^1_{N-k}$   
by  
\be    \label{funny1} 
\tilde   \rho_{k+1} (A_{k+1})   = \int   \de ( A_{k+1} - \cQ A_k)      \de ( \tau A_k  )  \rho_k(A_k)  \  DA_k   
\ee
Here   $  \de ( \tau A_k  ) $  is defined  as    in      (\ref{heathrow})  except that now 
 $y  \in   \bbT^{1}_{N-k}  $ and $x  \in \bbT^{0}_{N-k} $.    The other delta function is
 \be   
  \de (A_{k+1}  - \cQ   A_k   )=          \prod_{<y,y'>}    \de \B( (A_{k+1}  - \cQ   A_k   )(y,y') \B)   
\ee
Then we  define densities   
 $   \rho_{k+1}$     on fields  $A_{k+1}$ on    $ \bbT^0_{N-k-1}$        
by 
\be    
  \rho_{k+1}(A_{k+1}  )  =   L^{  \frac12 c_{k+1} }         \tilde  \rho_{k+1}  (A_{k+1,L})  \hs    c_k  =  (b_N- b_{N-k})   - (s_N -   s_{N-k})    
\ee  
Here in general for   a $d=3$ toroidal    lattice  with $L^M$ sites on side  we set   $b_M = 3L^{3M}$ as the number of bonds
and  $s_M  =  L^{3M}$ as the number of sites.

 Note that  if   $\rho_0$ is gauge invariant     then   $\rho_k$    gauge invariant for all $k$.  This follows by induction.   
 Assume it is true for  $k$   and  let  $\la$ be defined on  $\bbT^1_{N-k}$.    In   (\ref{funny1}) replace   $ \rho_k(A_k)$ 
 by  $  \rho_k(A_k  +  \pa  Q^T \la)$  and then compensate by the change of variables  $A_k \to  A_k -      \pa  Q^T \la$.
The  delta  function  $   \de ( \tau A_k  )$ is invariant since  $Q^T \la$ is constant on  the cubes  $B(y)$.   Using also
$\cQ   \pa  Q^T \la =  \pa  Q    Q^T \la  = \pa \la$  we   conclude 
  $\tilde   \rho_{k+1} (A_{k+1}- \pa \la)  =\tilde   \rho_{k+1} (A_{k+1}) $.  Hence  $\rho_{k+1}$ is gauge invariant as well.

\begin{prop}  For  $A_k$ on   $\bbT^0_{N-k}$  and      $\cA$  on       $\tk$ 
\be  \label{four} 
  \rho_k (A_k)  =      \int    \de (A_k -   \cQ_k  \cA )  \de^{\sX}_k   (\cA)    \rho_{0,L^{-k}} ( \cA) \   D\cA  \hs   1 \leq k  \leq  N-1    
\ee 
 In the  last  step we replace   $\de (A_k -   \cQ_k  \cA )$  by  $\de  (\cQ_N^{\bullet} \cA ) $ and have     
 \be  \label{fourN} 
  \rho_N   =      \int  \de  (\cQ_N^{\bullet} \cA )   \de^{\sX}_N   (\cA)    \rho_{0,L^{-N}} ( \cA) \   D\cA        
\ee
\end{prop}
\bigskip

\res
\begin{enumerate}
\item     
Here     $ \de^{\sX}_k   (\cA)$  is defined as before   
 \be  \de^{\sX}_k   (\cA)   =      \prod_{j=0}^{k-1}    \de  (\tau  \cQ_j \cA )   
 =     \prod_{j=0}^{k-1} 
 \prod_{ y_j,x_j     } 
 \de \B((\tau \cQ_j \cA)(y_j,x_j) \B)   
\ee
with   $    x_j  \in B(y_j),  x_j \neq  y_j $,
except that now   $y_j  \in   \bbT^{-k+j+1}_{N-k}  $ and $x_j  \in \bbT^{-k+j}_{N-k}$.   
\item  Since   $\rho_{0,L^{-N}} ( \cA) = f(\cA)\exp \left( - \frac12  \| \pa \cA \|^2 \right) $    
we  see that  $\rho_N$    is  the desired expression      (\ref{formal5}).
The basic  idea of  the renormalization group is  to control  the partition function  $\rho_N$   by controlling    the sequence  $\rho_0, \rho_1, \dots, \rho_N $.  
\item  Note  that  $ \cQ_k  \cA$   can  also be written with weighted sums  appropriate for the lattice  $\tk$.   For  oriented bonds 
we  have  
\be   
(  \cQ_k \cA) (y,  y +  e_{\mu} )  =  \int_{|x -y|  < \frac12  }      L^{-k}   \cA(  \Ga_{x,  x +   e_{\mu} } )    
\ee
where     $\int  [ \dots ] dx  =  \sum_x L^{-3k}   [ \dots ]  $  and   in general  
\be 
   L^{-k}   \cA (\Ga)   =     L^{-k} \sum_{b \in \Ga} A(b)  \equiv  \int_{\Ga} A
 \ee    
\end{enumerate} 
 \bigskip

\pr    In the next proposition we show that  $\rho_k$ is well-defined.  Assuming this  we  show that    the   representation   (\ref{four})
for  $\rho_k$   yield the representation  (\ref{four}) for   $\rho_{k+1}$.
We  have   
\be  \label{stun} 
\begin{split}
\tilde     \rho_k (A_{k+1})  = &  \int   \de ( A_{k+1} - \cQ A_k)      \de ( \tau A_k  )   \     \de (A_k -   \cQ_k  \cA )  \de^{\sX}_k   (\cA)    \rho_{0,L^{-k}} ( \cA) \   D\cA\   DA_k       \\
   = &  \int   \de ( A_{k+1} - \cQ_{k+1} \cA )      \de ( \tau   \cQ_k  \cA  )    \de^{\sX}_k   (\cA)    \rho_{0,L^{-k}} ( \cA) \   D\cA\        \\
   \end{split}
\ee
Replace  $A_{k+1}$  by  $A_{k+1,L}$ now with $A_{k+1}$ on  $\bbT^0_{N-k-1}$  and replace       $\cA$   by      $ \cA_L$    now  with  $\cA$  on  $\bbT^{-k-1}_{N-k-1}$. Use    the facts that     $\cQ$  is scale invariant,      $\cQ (\cA_L)  = ( \cQ \cA)_L$,   and that
$\tau$ is scale invariant,  $\tau  (\cA_L)  = ( \tau  \cA)_L$,  to obtain 
 \be  
\begin{split}
    \rho_k (A_{k+1})  
   = &  L^{  \frac12 c_{k+1} }       \int   \de \B(   (A_{k+1} - \cQ_{k+1} \cA)_L \B)      \de \B( (\tau   \cQ_k  \cA)_L  \B)    \de^{\sX}_k   (\cA_L)    \rho_{0,L^{-k}} (  \cA_L) \   D( \cA_L)\        \\ 
    \end{split}
\ee
However      $D( \cA_L)\     =     L^{ -\frac12  b_N}   D \cA $  and  
\be     
\begin{split}
 \de \B(   (A_{k+1} - \cQ_{k+1} \cA)_L \B)   = &    L^{ \frac12  b_{N-k-1} } \de \B(   A_{k+1} - \cQ_{k+1} \cA \B)  \\
   \end{split}
   \ee
Further     since   $\cQ_j \cA$ is a field on   $\bbT^{-k-1 +j}_{N-k-1}$  which has   $L^{N-j}$ sites on a side,
$ (\tau  \cQ_j \cA)$   takes  values  at    $s_{N-j} -  s_{N-j-1}$   points    and so    
\be   
\begin{split}
&  \de \B( (\tau   \cQ_k  \cA)_L  \B)    \de^{\sX}_k   (\cA_L)    
=          \prod_{j=0}^{k  }    \de  \B( (\tau  \cQ_j \cA)_L \B)    
=         \prod_{j=0}^{k  } L^{\frac12( s_{N-j} -  s_{N-j-1})}       \de  \B( \tau  \cQ_j \cA \B)     
=    
      L^{\frac12(s_N  -  s_{N-k-1})}     \de^X_{k+1} (\cA )    \\
\end{split}
\ee
The powers of  $L$ from under the integral sign  collect to form   $L^{-\frac12c_{k+1}}$ which    cancels  the 
 $L^{\frac12c_{k+1}}$  in front.   Thus we have the desired
\be  
\begin{split}
     \rho_{k+1} (A_{k+1})  
     =   &     \int   \de ( A_{k+1} - \cQ_{k+1} \cA )        \de^{\sX}_{k+1}   (\cA)    \rho_{0,L^{-k-1}} (  \cA) \   D \cA\        \\ 
      \end{split}
\ee
This completes  the proof
\bigskip

\begin{prop} \label{well}    If   $F_0$  is exponentially bounded  $ \rho_k $   is well-defined for  $1 \leq  k  \leq N$.
\end{prop}
\bigskip

\pr  (after   \cite{BaJa86},   \cite{BIJ85}  for  $k=1$, $\tau^0$.)   
For   $k<N$ we have   
\be  
  \rho_k (A_k)  = \int       \de (A_k -   \cQ_k  \cA )  \de^{\sX}_k   (\cA)  F_{0, L^{-k}} (\cA)   \exp   \B(  -  \frac12  \| d \cA \|^2   \B)  D\cA
\ee
The delta functions restrict the  integral to the surface   
\be    \label{constrain}
   \cQ_k  \cA =A_k,  \  \ \tau \cQ_{k-1}  \cA  = 0,\  \ \dots,\  \ \tau \cQ \cA  =0, \  \ \tau \cA  =0
\ee
We first note that  this surface is not empty.    There are certainly  $\cA$ satisfying the null conditions since they
represent the kernel of a linear operator  to a lower  dimension space.   These conditions do not  involve  bonds    
joining  neighboring unit cubes     $B^k(y),  B^k(y')$  in  $\tk$.    By adjusting   $\cA(b)$ for such bonds  there is plenty 
of freedom to  meet the final    condition     $ \cQ_k  \cA =A_k$.

 Let  $\cA_0$ satisfy    (\ref{constrain}).   Change variables   by     $\cA  = \cA_0 + \cZ$
and then     the integral  is  
\be  \label{sunny} 
  \rho_k (A_k)  = \int       \de (  \cQ_k  Z )  \de^{\sX}_k   (\cZ) F_{0, L^{-k}} (\cA_0 +  \cZ  ) 
    \exp   \B(  -  \frac12  \| d \cA_0 \|^2    -  \frac12  \| d \cZ \|^2  -   <d\cA_0, d \cZ>   \B)  D\cZ
\ee
It   suffices to show  for  $\cZ \in \tk$     that    $  \| d \cZ \|^2  $ is positive definite   on the  surface  
\be    \label{constrain2}
   \cQ_k  \cZ =0,  \  \ \tau \cQ_{k-1}  \cZ  = 0,\  \ \dots,\  \ \tau \cQ \cZ  =0, \  \ \tau \cZ  =0
\ee
 Then this integral converges and the original integral converges.

First suppose   $k=1$ so  we need to show  for  $\cZ \in \bbT^{-1}_{N-1}$     that  $\cQ \cZ = 0$ and  $\tau \cZ  =0$ and  $d  \cZ =0$ imply   $\cZ =0$.  
Arguing as in Proposition   \ref{santora},  but with the $L$-cube  $T_0$ replaced by a unit cube  $T_y$ centered on  $y \in  \bbT^0_{N-1}$,  we find that  
$\cZ$ vanishes  on each  cube  $B(y)$. 
  Consider   $\cZ(b)  $  for bonds  connecting  neighboring cubes $B(y), B(y')$.   The condition
$d \cZ =0$  implies that they all have the same  value.  Then  the condition $(\cQ \cZ)(y,y') = 0$   implies that that value is 
zero.

Back to the general case we      note that  $d \cZ = 0$ implies   $d \cQ  \cZ  =0$.  This follows since   if   
$x \in \bbT^{-k}_{N-k}$, $y \in \bbT^{-k+1}_{N-k}    $  and $ \eta = L^{-k}$:
\be 
\begin{split}
&  (d \cQ  \cZ) \B(  [ y,  y + L \eta  e_{\mu} ,  y + L \eta  e_{\mu}+  L \eta  e_{\nu}, y + L \eta  e_{\nu},y] \B) \\
  = &\sum_{x \in B(y) }   \cZ   \B(\Ga  [ x,  x + L \eta  e_{\mu} ,  x + L \eta  e_{\mu}+  L \eta  e_{\nu}, x + L \eta  e_{\nu},x] \B) \\
\end{split}
\ee
where   $\Ga[ \cdots  ]$  in the closed path      passing through the indicated points.   But     $\cZ  (\Ga[ \cdots  ])=0 $
follows from  $d \cZ =0$ and the lattice Stokes theorem.   Hence the result.  More generally     $d \cZ = 0$ implies   $d \cQ_j  \cZ  =0$
for any  $j$.
  
Now for   $k< N$    we argue that      $\cQ_{k} \cZ =  \cQ (\cQ_{k-1}\cZ) =   0$ and  $\tau  \cQ_{k-1} \cZ  =0$ and  $d \cQ_{k-1} \cZ =0$
  imply   $\cQ_{k-1} \cZ  =0  $ just as for  $k=1$.   Then    $\cQ_{k-1} \cZ =\cQ  ( \cQ_{k-2} \cZ)=   0$ and  $\tau  \cQ_{k-2} \cZ  =0$ and  $d  \cQ_{k-2}  \cZ =0$
  imply   $\cQ_{k-2} \cZ  =0  $.  Continuing  we reduce to the case  $k=1$ and the conclusion  $\cZ  $ = 0.

For  $k=N$ we have  $A_N=0$ and    are looking at   the same  expression   (\ref{sunny}),  but with   $\de (  \cQ_k  Z )$ replaced with
$ \de  (\cQ_N^{\bullet} \cZ ) $.   
  Then    $\cQ^{\bullet}_N \cZ  = \cQ^{\bullet} (  \cQ_{N-1}  \cZ  ) =0$  and    $\tau  \cQ_{N-1} \cZ =0 $ and   $d  \cQ_{N-1} \cZ =0$
  imply  the same with   $\tau^0  \cQ_{N-1} \cZ =0 $  and hence    $\cQ_{N-1} \cZ =0 $  on all bonds as in section \ref{torus}.   The rest of the argument proceeds as before.

Thus  in all cases    $ \| d \cZ \|^2   \geq   \const   \|\cZ \|^2  $.   The constant is not    uniform  in  $N$,   but the bound is  
sufficient to show    the integral is well-defined.  This completes the proof.
   \bigskip

\subsection{minimizers}

Let   $\cH^{\sx}_{ k}A_k  $  be the minimizer of    $ \| d \cA  \|^2 $  on the subspace  (\ref{constrain}). 
One    can obtain    explicit   representation of  $\cH^{\sx}_k$, see   \cite{BIJ85}, \cite{Imb86}.  This   representation
involves    a certain   Green's  function  $G^{\sx}_k$ which is essentially   the inverse   of the operator  defined by the
quadratic form   $\| d \cZ \|^2 $ on the   surface (\ref{constrain2})
  with $\tau^0$ instead  of  $\tau$.
The same representation holds for our case if we define  $G^{\sx}_k$   as the inverse
on the   surface (\ref{constrain2}).   In any case  for this paper we    do not need to know much about
$\cH^{\sx}_{ k}$ beyond existence.  It is not very regular and is not  used directly.

Expand around the minimizer    as  above   by    $A  = \cH^{\sx}_k  A_k  + \cZ$.   The linear term vanishes     we find  
\be  \label{sex} 
      \rho_k(A_{k})   =    \sZ_{k}       F_k (\cH^{\sx}_kA_k)     \exp \B(   - \frac12 <A_k,  \De_k  A_k >   \B)   
 \ee
where
\be   \label{military}
\begin{split}
    <A_k,  \De_k  A_k >   = &    \|  d \cH^{\sx}_{k} A_k  \|^2  \\    
 F_k (\cH^{\sx}_kA_k )   
   = & \sZ_k^{-1}    \int    \de (  \cQ_k \cZ  ) \de^{\sX}_k   (\cZ)  F_{0,L^{-k}}  (\cH^{\sx}_kA_k   + \cZ)    \exp   \B( - \frac 12 \|  d \cZ \|^2    \B)  D \cZ
    \\   
\sZ_{k}   =   &   \int    \de (  \cQ_k \cZ  )  \de^{\sX}_k   (\cZ)   \exp  \B ( - \frac 12 \|  d \cZ \|^2   \B )  D \cZ \\
\end{split}
\ee
The   $F_k$  are     functions  of  fields    defined on bonds in $\tk$. 
      Controlling  the sequence  $F_1, F_2,  \dots  $ is the main issue for a 
complete analysis.   To study this we consider how to pass from  $F_k$ to $F_{k+1}$.

\subsection{the next  step}

Suppose we are starting with the expression  (\ref{sex}).   In the next step  we  consider 
\be  \label{six}
\begin{split}
\tilde \rho_{k+1} (A_{k+1})  = &      \int    \de (A_{k+1} -    \cQ A_k  )\   \de(  \tau  A_k)  \    \rho_k(A_k)    DA_k \\
 =   & \sZ_{k}      \int    \de (A_{k+1} -    \cQ A_k  )\   \de(  \tau  A_k)  \     F_k (\cH^{\sx}_k   A_k)     \exp \B(   - \frac12 <A_k,  \De_k  A_k >   \B)   
 DA_k  \\
 \end{split}
\ee
Let   $ H^{\sx}_{k} A_{k+1}$  be  the minimizer for   $   <A_k, \De_k  A_k  >  $  subject to the constraints.   Again
an explicit expression can be found     \cite{BIJ85}, \cite{Imb86}.    Expanding around the minimizer  with   $A_k =  H^{\sx} _{k} A_{k+1} + Z$  
we   get  
\be     \label{spotless}
\begin{split}
\tilde \rho_{k+1} (A_{k+1})=& \sZ_{k} \sZ^f_k \exp \B(   - \frac12<   H^{\sx}_{k} A_{k+1} ,  \De_{k}  H^{\sx}_{k} A_{k+1}   >  \B)    F^*_{k}  (\cH^{\sx}_kH^{\sx}_k  A_{k+1}  ) \\
\end{split}   
\ee
where  
\be    \label{sammy}
\begin{split}
    F^* _{k}  (\cH^{\sx}_kH^{\sx}_k  A_{k+1}   ) 
   = & (\sZ^f_k)^{-1}    \int     \de (  \cQ   Z  )   \de(  \tau Z)  \    F_k \B( \cH^{\sx}_kH^{\sx}_k  A_{k+1}     + \cH^{\sx}_kZ\B)     \exp \B(   - \frac12 <Z,  \De_k Z >   \B)  D Z
    \\   
\sZ^f_{k}   =   &   \int    \de (  \cQ   Z  )   \de(  \tau Z)  \       \exp \B(   - \frac12 <Z,  \De_k Z >   \B)  D Z \\
\end{split}
\ee
After  scaling then we  have     
\be     \label{spotless2}
\begin{split}
 \rho_{k+1} (A_{k+1})=& \sZ_{k} \sZ^f_k L^{\frac12 c_{k+1}}  \exp \B(   - \frac12<   H^{\sx}_{k} A_{k+1,L} ,  \De_{k}  H^{\sx}_{k} A_{k+1,L}   >  \B)     F^*_{k}  (\cH^{\sx}_kH^{\sx}_k  A_{k+1,L}  )  \\
\end{split}   
\ee

Take the special case in  which $F_0(A_0) = 1$. Then    $F_k =1$ and $F^*_k=1$     for all $k$.   Then taking      $A_{k+1} =0$  and comparing   (\ref{spotless2})
with   (\ref{sex})  for   $k+1$   we  have the identity 
\be 
   \sZ_{k+1}  =  \sZ_k \sZ_k^f    L^{\frac12 c_{k+1}} 
    \ee
We  also  see that the quadratic form  in    (\ref{spotless2})  must  be   $  < A_{k+1},  \De_{k+1}  A_{k+1} >$.

Now in the general  case   (\ref{spotless2}) says       
\be  
  \rho_{k+1} (A_{k+1})= \sZ_{k+1} \exp \B(   - \frac12< A_{k+1},  \De_{k+1}  A_{k+1} >  \B)    F^*_{k}  (\cH^{\sx}_kH^{\sx}_k  A_{k+1,L}  )  
\ee
Comparing this  with     (\ref{sex})  for   $k+1$    yields
\be     \label{pinky} 
   F^*_{k}  (\cH^{\sx}_kH^{\sx}_k  A_{k+1,L}  )  
=    F_{k+1} ( \cH^{\sx}_{k+1}  A_{k+1}  )       \ee

Next      take the special case in  which $F_0(A_0) = <A_0,J>$.   It is not gauge invariant, but does not have to be for this argument. 
From (\ref{military}) we  get  $F_k( \cH^{\sx}_k A_k) =  < (\cH^{\sx}_k A_k)_{L^k}, J>$ for all $k$.   Then    (\ref{sammy}) yields
 $F^*_k( \cH^{\sx}_k H^{\sx}_kA_{k+1}) =  < (\cH_k  H^{\sx}_kA_{k+1})_{L^k}, J>$ for all $k$.    Hence      (\ref{pinky}) says     
\be     
<  (\cH^{\sx}_kH^{\sx}_k  A_{k+1,L}  )_{L^k}, J>
=<(\cH^{\sx}_{k+1}  A_{k+1}  )_{L^{k+1}} ,J>        \ee
This yields  the identity
\be  
 \cH^{\sx}_{k}   H^{\sx}_{k} A_{k+1,L}   
=( \cH^{\sx}_{k+1}A_{k+1})_L 
 \ee

Back to    the general case (\ref{sammy})  with  $A_{k+1}  \to  A_{k+1,L}$    becomes       
  \be   
     F_{k+1}  ( \cH^{\sx}_{k+1}A_{k+1} )   
   =    (\sZ^f_k)^{-1}    \int      \de (  \cQ   Z  )   \de(  \tau Z)  \    F_k \B( ( \cH^{\sx}_{k+1}A_{k+1} )_L       + \cH^{\sx}_k Z  \B)    \exp \B(   - \frac12 <Z,  \De_k Z >   \B)  D Z
\ee 
More generally we  define \textit{fluctuation integrals}  for  any       $\cA$ on  $\bbT^{-k-1}_{N-k-1}$    by
 \be   \label{owen}
     F_{k+1}  ( \cA )   
 =    (\sZ^f_k)^{-1}    \int     \de (  \cQ   Z  )   \de(  \tau Z)  \    F_k \B( \cA _L       + \cH^{\sx}_k Z\B)    
  \exp \B(   - \frac12 <Z,  \De_k Z >   \B)  D Z
\ee  
Here $Z$ is on $\bbT_{N-k}^0$  while    $\cH^{\sx} Z$  and $\cA_L$   are   on   $\tk$. 
This generates the sequence  $F, F_1,F_2, \dots$.    If  $F_0$ is gauge invariant then  $F_k$ is gauge invariant for all $k$.

\subsection{Feynman  gauges}

Consider   integrals    of the form  
and
\be  \label{fourfour} 
  \rho_k (A_k)  =      \int    \de (A_k -   \cQ_k  \cA ) \de^{\sX}_k   (\cA) f(\cA) \exp \B( - \frac12 \| d \cA \|^2  \B)   D\cA  
  \ee
 where  $f(\cA) $ is a gauge invariant function  on  fields  $\cA$ defined on  $\tk$. 
  The expression    $  (\ref{four}) $  is of this form.   
  Also let   $\de=d^T$   be  the adjoint of $d = \pa$ on scalars   ($\de$ is  the divergence),  and  let     $R_k $  be  the projection onto the subspace      $\De ( \ker \   Q_k  )  $.  
 
  We    introduce    the  modified Feynman gauge   developed   in   \cite{Bal84a},  \cite{BaJa86}:   
  
\begin{prop}     The integral (\ref{fourfour})     can be expressed     for  any     $\al >0$  as      
\be       \label{int5} 
 \rho_k (A_k)    =  \const       \int      \de (A_{k  } -    \cQ_k \cA  )\  \ 
   f(\cA)       \exp \B(  -  \frac{1}{2 }     \|  d  \cA \|^2  -   \frac{1}{2 \al} < \de \cA,  R_k\   \de \cA>  \B)     D\cA        
\ee
This  includes  the Landau gauge at   $\al =  0$  in which case 
\be       \label{int6} 
 \rho_k (A_k)    = \const       \int      \de (A_{k  } -    \cQ_k \cA  )    \de_{R_k} ( R_k  \de  \cA    )   \ 
   f(\cA)       \exp \B(  -  \frac{1}{2 }     \|  d  \cA \|^2   \B)        D\cA        
\ee
where  $ \de_{R_k}  $ is the delta function in the subspace  $\ran\ R_k   = \De ( \ker \   Q_k  )  $.  
\end{prop}  
\bigskip

\pr  We sketch the proof.    One employs    a Fadeev- Popov procedure.
Define
\be
  \sZ_k (  \cA)  =     \int    \de (Q_k \la)   \exp   \B(  - \frac{1}{2 \al}  \| \de \cA^{\la}  \|^2  \B)  \ D  \la
\ee     
and   insert   $1=  \sZ_k (  \cA)  / \sZ_k (  \cA)  $
under the integral sign in (\ref{fourfour}).   In the numerator  change the order of integration.  This yields 
\be       \label{int2} 
 \int \left[  \int     \sZ_k (  \cA)^{-1}   \de (A_{k } -    \cQ_k  \cA  )\   \de^{\sX}_k   (\cA)  \  f(\cA)   \exp \B(  - \frac12    \|  d  \cA \|^2    - \frac{1}{2 \al}   \| \de \cA^{\la}   \|^2  \B)\     D\cA  \right]        \de (Q_k \la) \   D \la
\ee
Now  let   $\cA \to  \cA^{-\la} = \cA +  \pa  \la$.   The  delta function   $  \de (A_{k  } -    Q_k \cA  )$ is invariant since
$   \cQ_k  \pa \la   =   \pa Q_k \la    =0$    and       $ \|  d  \cA \|^2 $
is invariant.   Furthermore  $\sZ_k (  \cA)$   is  invariant since $Q_k \la = 0$  implies
\be  
\begin{split}
  \sZ_k (  \cA^{-\la}) =  &    \int     \de (Q_k \la')   \exp   \B(  -  \frac{1}{2 \al}   \| \de \cA^{\la'- \la}  \|^2  \B) D \la' \\
=  &    \int     \de (Q_k    \la'   )  \exp   \B(  - \frac{1}{2 \al}   \| \de \cA^{\la'}  \|^2  \B)\ D \la'   =   \sZ_k (  \cA)       \\ 
\end{split}
\ee
Thus our expression becomes  
\be       \label{int3} 
 \int \left[  \int     \sZ_k (  \cA)^{-1}   \de (A_k   -    \cQ_k \cA   )\   \de^{\sX}_k(\cA^{-\la} )  \   f(\cA)
   \exp \B(  - \frac12   \|  d  \cA \|^2   -  \frac{1}{2 \al}   \| \de \cA   \|^2  \B)\     D \cA  \right]        \de (Q_k \la)   D \la
\ee
Now  change the order of integration again.   The $\la$ integral is   $  \int  \de^{\sX}_k  ( \cA^{-\la})  \         \de (Q_k \la)   D  \la  $   
and as    in Proposition  \ref{three}  it is constant.  
Thus     
\be       \label{int4}    \rho_k (A_k)  = \const      \int     \sZ_k (  \cA )^{-1}   \de (A_{k  } -    \cQ_k \cA  )\  \   f(\cA)  \exp \B(  -  \frac{1}{2 }     \|  d  \cA \|^2  -   \frac{1}{2 \al}  \| \de \cA   \|^2  \B)\     D\cA        
\ee
Finally  a computation    \cite{Bal84a} shows that  
\be 
    \sZ_k (  \cA)  =   \const    \exp  \B(  - \frac{1}{2 \al}   <   \de \cA,  P_k \   \de  \cA   >   \B)  
     \ee 
     where  
  \be  P_k =  G_k Q_k^T (Q_k G_k^2 Q_k^T)^{-1} Q_k G_k
  \ee
  and   
  for any  $a \geq 0$   
  \be  \label{gk}
  G_k  =  ( -    \De + a Q_k^T Q_k  ) ^{-1}
  \ee      
  This gives the Feynman gauge expression (\ref{int5})   with  $R_k = I  - P_k$.   One verifies that  $R_k$ is the projection on 
  $\De ( \ker \   Q_k  )  $ as  claimed.
  
    The Landau gauge expression   (\ref{int6})   follows
by taking the limit  $\al \to   0$.    Alternatively one  can  use the Fadeev- Popov procedure again to pass from  
from  (\ref{int5}) to (\ref{int6});  see the appendix.   This completes the proof. 
\bigskip

Let   $\cH_kA_k $  be the minimizer  of    $ \frac12 \| d \cA  \|^2 -   (2 \al)^{-1}  <   \de \cA,  R_k    \   \de  \cA   > $    subject to the constraint  
$\cQ_k \cA  = A_k$   imposed    in  (\ref{int5}).   Then   one can establish the following facts concerning this Feynman gauge minimizer:

\begin{enumerate}
\item   $\cH_k$ is independent of $\al$ and is also the minimizer for  Landau gauge.  \cite{BaJa86}
\item    $ \cH^{\sx}_{k}  =    \cH_{k}  +   d  D_k$     for some operator  $D_k$. \cite{BIJ85}, \cite{Imb86}
\item   $\cH_k,  \pa \cH_k,  \de_{\al}  \pa  \cH_k  $ 
have kernels with exponential decay.   ($\de_{\al}$ is the Holder derivative of order $\al$.) \cite{Bal84a}, \cite{Bal84b}.
\end{enumerate}    

Point  (2.)  says that in  a gauge invariant expression we 
can replace  the axial gauge minimizer  $ \cH^{\sx}_{k} $ by    $ \cH_{k} $.   Point  (3.) says this is useful  since     $\cH_k$  has good regularity  and decay 
properties.   Note in particular that     
instead of     $ <A_k,  \De_k  A_k >   =   \|   d  \cH^{\sx}_k A_k \|^2  $
we can  take  
\be     \label{sizemore}
 <A_k,  \De_k  A_k >   =   \|   d  \cH_k A_k \|^2  
\ee

\subsection{a lower bound on  $\De_k$}

For the quadratic form  $\De_k$ we have a lower bound independent of   $N,k$.  

\begin{prop}  
 There is a constant  $C$  (depending on $ L$)  such that  for $A$  on  $\tz$  and satisfying  
      $ \cQ A  =0, \tau A  =0  $
\be    \label{blue}
<A,  \De_k A>\   \geq\   C \| A  \|^2
\ee
\end{prop}
\bigskip

\pr  
The proof     follows       lemma  2.4  in  \cite{Bal84b},   with minor modifications for the  covariant axial gauge
and specialized to $d=3$. 
 Start with   (\ref{sizemore}).
Using an  explicit formula for  $\cH_k$  and working in Fourier  transform space one shows  (section D in  \cite{Bal84a} ) 
\be
<A,  \De_k A>\   \geq\   C \| d  A  \|^2
\ee
Thus   our  claim is reduced to showing  $ \| d  A  \|^2  \geq\   C \| A  \|^2$,    and since  $\cQ A =0$  this is equivalent to 
showing  that  for   $\tau A  =0$
\be   \label{sump}
  \| d  A  \|^2   +   \| \cQ A\|^2      \geq\   C \| A  \|^2
\ee
The proof divides into three parts 
\begin{enumerate}
\item  We  first  consider  an $L$-cube  $B(y)$ centered on  $y \in \bbT^1_{N-k}$ and show     
 that   (c.f   (2.123)  in  \cite{Bal84b}) 
\be   \label{thing1}
 \| A  \|^2_{B(y) }  \leq       3L^3    \| d  A  \|^2_{B(y) }      \ee
To  see this  note  that   for any  permutations  of coordinates  $\pi$  and any bond  $<x, x+ e_{\mu}>$  in  $B(y)$
\be  A(  \Ga^{\pi} _{yx} )   +    A(x, x+e_{\mu} )    -  A(\Ga^{\pi}_{y, x + e_{\mu} })
\ee      
is  a closed curve.   Hence  it  bounds a surface  $\Si^{\pi}_{y,x,\mu}$ and by Stokes theorem    the expression is equal to 
 $dA (\Si^{\pi}_{y,x,\mu})$.
Averaging  over   permutations  gives
\be    (\tau A )(y,x)    +  A(x,  x+ e_{\mu})    -  (\tau A)(y, x + e_{\mu}  )   =  \frac{1}{d!}   \sum_{\pi}   dA (\Si^{\pi}_{y,x,\mu}) 
\ee
But   $\tau A  =0$   and so    
\be     A(x,  x+ e_{\mu})  =  \frac{1}{d!}   \sum_{\pi}   dA (\Si^{\pi}_{y,x,\mu}) 
\ee 
Now 
\be|  dA (\Si^{\pi}_{y,x,\mu}) |     \leq     \sum_{p  \in      \Si^{\pi}_{y,x,\mu} }   |dA(p) | 
 \leq    \B(\sum_{p  \in      \Si^{\pi}_{y,x,\mu} }  1  \B)^{\frac12}
\B(    \sum_{p  \in      \Si^{\pi}_{y,x,\mu} }   |dA(p)|^2  \B)^{\frac12 }  \leq   L^{\frac12} \B(    \sum_{p  \in      \Si^{\pi}_{y,x,\mu} }   |dA(p)|^2  \B)^{\frac12 }  
  \ee
  and      
\be 
\begin{split}
& \sum_{<x, x+ e_{\mu}>\in B(y)}   | dA (\Si^{\pi}_{y,x, \mu})|^2\   \leq\    L  \sum_{<x, x+ e_{\mu}> \in B(y) }  
 \sum_{p  \in      \Si^{\pi}_{y,x,\mu} }   |dA(p)|^2   \\
 & =    L   \sum_{p  \in B(y)   }   |dA(p)|^2  \sum_{<x, x+ e_{\mu}>:    \Si^{\pi}_{y,x,\mu} \ni p}  1\  \leq \  3L^3  \| dA \|^2_{B(y)} \\
 \end{split}
\ee
This yields
\be  \| A  \|_{B(y)}  \leq     \frac{1}{d!}   \sum_{\pi}      \| dA (\Si^{\pi}_{y, \cdot } ) \|_{B(y)}    \leq 3^{\frac12}  L^{\frac32}    \| dA \|_{B(y)} 
\ee
which    proves     (\ref{thing1})

\item   Now consider  bonds or  plaquettes   that  join neighboring unit cubes  $B(y), B(y')$  denoted  $B(y,y')$.    One  shows that  (  (2.127)  in  \cite{Bal84b}  )  \be 
\begin{split}   
&   \sum_{ p   \in    B(y,y') } | dA (p) |^2    + |\cQ    A (y,y')  |^2  \\
\geq   &\   \frac{1}{3 L^4}        \sum_{b  \in B(y,y')}  |A(b)|^2    
-  \frac{1}{ L^3}  \sum_{b \in B(y) }  |A(b)|^2   -   \frac{1}{ L^3}  \sum_{b \in B(y') }  |A(b)|^2  \\
\end{split} 
\ee
Summing  over  oriented bonds   $<y,y'>$    and using that    
\be
  \sum_{<y,y'>}    
  \sum_{b \in B(y) }  |A(b)|^2   =  3   \sum_y \sum_{b  \in B(y)}     |A(b)|^2 
\ee
we have     
\be  \label{thing2}
\begin{split}
  &   \sum_{<y,y'>}  \B(     \sum_{ p   \in    B(y,y') } | dA (p) |^2  +   |\cQ    A (y,y')  |^2 \B)\\
\geq      &     \   \frac{1}{3 L^4}     \sum_{<y,y'>}       \sum_{b  \in B(y,y')}  |A(b)|^2    -
\frac{6}{ L^3} \sum_y \sum_{b \in B(y) }  |A(b)|^2   \\  
   \end{split}
\ee

\item
We  combine   (\ref{thing1}) and (\ref{thing2}) to     estimate   \be
\begin{split}
& \|  dA  \|^2   +     \|\cQ    A    \|^2    \\
=  &   \sum_y   \sum_{p \in B(y)} | dA (p) |^2    + \sum_{<y,y'>}  \B(     \sum_{ p   \in    B(y,y') } | dA (p) |^2  +   |\cQ    A (y,y')  |^2 \B)\\
\geq    &
  \frac{1}{3L^3}  \sum_y   \sum_{b \in B(y)  }    |A(b)|^2
+   \frac{1}{36}    \B(   \frac{1}{3 L^4}       \sum_{<y,y'>}       \sum_{b  \in B(y,y')}  |A(b)|^2    -
 \frac{6}{ L^3} \sum_y \sum_{b \in B(y) }  |A(b)|^2   \B)   \\
=
  &  \frac{1}{6L^3} \sum_y   \sum_{b \in B(y)} | A (b) |^2      
      +   \   \frac{1}{108L^4 }   \sum_{<y,y'>}       \sum_{b  \in B(y,y')}  |A(b)|^2       
\geq       \frac{1}{108L^4 }      \| A\|^2  \\
  \end{split}
\ee
Thus  (\ref{sump}) is established and the proof is complete.   
\end{enumerate}

\newpage

\subsection{parametrization of  the   fluctuation integral} 
\label{parametrization}  
Next we   parametrize the  fluctuation integrals (\ref{owen}),   or more generally  integrals of the form  
\be  \label{money}
    \int    \de(  \tau Z)   \de (  \cQ   Z  )  \   
    f(  Z)      \exp \B(   - \frac12 <Z,  \De_k Z >   \B)  D Z \Big/  \{ f =1 \}   
\ee
where  $ \{ f =1 \}  $  is  the same integral with  $f=1$.  
The analysis is a variation of    \cite{Bal84b}.

The field $Z$ is a function on bonds  in a   unit lattice like $\tz$.   
We  split the bonds into  those in $L$-cubes  and those  joining  $L$-cubes  
by  $Z = (Z_1,Z_2) $  where  
\be
\begin{split}
Z_1   =&   \{  Z(b)  \}   \hs    b \in \bigcup_y  B(y)   \\
Z_2   =&   \{  Z(b)  \}   \hs    b \in \bigcup_{<y,y'>}  B(y,y')   \\
\end{split}
\ee
The integral over  $  \de(  \tau Z)    =   \de(  \tau Z_1) $   is just  an integral over the subspace    
   $\ker  \tau $ as in  section  (\ref{this}).     Thus  with  $\tilde Z_1  \in \ker \tau$ we have    
\be
  \label{money2}
    \int       
 \B[     \de (  \cQ   Z  )   f(  Z)      \exp \B(   - \frac12 <Z,  \De_k Z >   \B)  \B]_{Z = ( \tilde Z_1, Z_2 ) }    D \tilde Z_1    D Z_2  \Big/  \{ f =1 \}   
\ee
For the    remaining  delta function       we   are    integrating over  the  subspace      
$ \cQ  Z =  \cQ( \tilde Z_1, Z_2)  =0 $.    Let  $b(y,y')$
be the central bond in $B(y,y')$    and let        $\tilde Z_2$ be the non-central bonds:
\be
 \tilde    Z_2   =      \{  Z(b)  \}   \hs    b \in \bigcup_{<y,y'>} \B( B(y,y')  - b(y,y')  \B)   
\ee 
 The condition  is now     $ \cQ\B( \tilde  Z_1,  \tilde  Z_2,  \B  \{ Z( b(y,y'))\B  \} \B)    =0$ and    can be solved for   the   variables      $Z( b(y,y')  )$  and written as
\be
Z( b(y,y')  )  =   \B(  S( \tilde Z_1 ,  \tilde Z_2) \B) ( b(y,y')  )
   \ee   
for some local   linear operator  $S$.   For an explicit formula for  $S$ see  \cite{Dim14}.     Then      $Z_2  = ( \tilde Z_2,   S ( \tilde Z_1,  \tilde Z_2  ) ) $   and   we  can evaluate the delta function   by      
  \be
  \label{money3}
    \int       
 \B[      f(  Z)      \exp \B(   - \frac12 <Z,  \De_k Z >   \B)  \B]_{Z = ( \tilde Z_1,  \tilde  Z_2, S( \tilde Z_1, \tilde Z_2  ) ) }  
   D \tilde Z_1    D \tilde    Z_2  \Big/  \{ f =1 \}   
\ee
Finally   put      $ \tilde  Z    = ( \tilde  Z_1,\tilde     Z_2)$     and define    
\be 
 Z    =      C \tilde Z   = \big(  \tilde Z,  S( \tilde Z   )  \big)  
\ee
Note 
that  $C$ is a local operator mapping  to      the  subspace  $\cQ Z  =0,  \tau Z =0$.      
Now  (\ref{money3})   can be written   
\be   \label{money4}
     \int       
       f( C  \tilde    Z)      \exp \B(   - \frac12 <C \tilde Z,   \De_k   C \tilde Z   >   \B)    D \tilde Z    \Big/  \{ f =1 \}   
\ee
Finally define      
\be     C_k   =   ( C^{ T } \De_k  C  )^{-1}     
\ee
and identify the Gaussian  measure $\mu_{C_k}$  with covariance   $C_k$.   Then (\ref{money4}) and hence   (\ref{money})
is expressed as    
\be   \label{money5}
    \int               f( C  \tilde    Z)   d \mu_{C_k }  ( \tilde Z  )     
\ee

\subsection{representation for  $C_k$}

To analyze integrals  like  (\ref{money5}) note that  $C^T\De_kC$ is a  uniformly  bounded strictly positive operator with     exponentially decaying kernel
 by  (\ref{sizemore}), (\ref{blue}).    It follows   by a lemma of Balaban  (\cite{Bal83b}, section 5)    that  
 $C_k =(C^T\De_kC)^{-1}$ has the same properties .  Then one  can employ a cluster expansion to get estimates
on the integral.   However  we eventually want to use a  version of the cluster expansion which employs  a random walk 
expansion for  $C_k$. For this Balaban's lemma is not sufficient.   The analysis     is not straightforward   since   $C_k$ is not the inverse of a local operator.
   
  We   develop another  representation for     $C_k$.    
 The following is a simpler version of an analysis by Balaban in    \cite{Bal85b}.   There he treats a multi-scale non-abelian problem
 while here  it is single scale and abelian.  It is easier to   consider          $CC_kC^T$  and  the argument  has    a number of steps.   
\bigskip

 \noindent  \textbf{step 1:}
 Start with the representation  for  $J, Z$ on  $\tz$
\be  \label{dogged} 
\begin{split}
\exp \B( \frac 12  <J,   CC_k C^TJ>  \B )       
 =   &   \int    e^{   <C \tilde  Z , J>  }
   \exp \B(  - \frac12   <C  \tilde  Z , \De_k C  \tilde Z   >   \B)    \    D  \tilde Z  \B/  \{ J=0\}   \\
  =   &    \int   e^{  <     Z , J>  } 
      \de (  \cQ   Z  )   \de(  \tau Z)  \     \exp \B(   - \frac12 <Z,  \De_k Z >   \B)  D Z \B/  \{ J=0\} 
  \\
\end{split}
\ee   
Combining  (\ref{sex}) and      (\ref{int6} )  with $f=1$
we have
\be       \exp \B(   - \frac12 <Z,  \De_k Z >   \B)  
=   \const   \int    \de (Z -   \cQ_k  \cA ) \de_{R_k}( R_k \de \cA)   \exp \B( - \frac12\| d \cA \|^2 \B)    D\cA 
\ee  
and we insert this in (\ref{dogged}).
It is tempting to  now   do the integral over  $Z$.  But this turns out to   lead to difficulties    so we postpone it.  
Instead we further complicate things by inserting  for  $\la$ on $\tk$ 
\be 
  1 = \const   \int  \  \de (Q_{k+1} \la )  \de_{R_{k+1}}\B( R_{k+1}( \de \cA - \De \la)  \B)\  D \la
\ee
See Proposition  \ref{lunk}  in the appendix for details.   This will free up some intermediate gauge fixing.    
Then  (\ref{dogged}) becomes  
\be  \label{dogged2} 
\begin{split}
   &      \const   \int        \de (  \cQ   Z  )  \de (Z -   \cQ_k  \cA )\   \de(  \tau Z)  \de_{R_k}\B( R_k \de \cA\B) 
    \de_{R_{k+1}}\B( R_{k+1} ( \de \cA - \De \la )  \B)  \de (Q_{k+1} \la )\\
&  \exp \B( - \frac12\| d \cA \|^2 +<     Z , J> \B)\    D\cA\  D Z\  D \la  \\
  \\
\end{split}
\ee

\noindent  \textbf{step 2:}
Next let  $\cA  \to \cA^{\la} = \cA -  \pa  \la $  so $\de \cA  \to   \de \cA + \De \la$.  We   also write  for  $\mu$ on  $\bbT^0_{N-k}$
\be  
  \de (Q_{k+1} \la )   =  \int  \de(  Q\mu  )  \de  (   \mu - Q_k \la)    D \mu
 \ee
 and use   $\cQ_k \pa  = \pa Q_k$ 
to obtain   
\be  \label{dogged3} 
\begin{split}
  &      \const    \int        \de (  \cQ   Z  )  \de (Z -   \cQ_k  \cA  -   \pa  Q_k \la )\ 
  \de(  \tau Z)  \de_{R_k}\B( R_k  ( \de \cA +  \De \la  ) \B)  \de_{R_{k+1}}\B( R_{k+1} \de \cA  \B)  \\
& \de(  Q\mu  )  \de  (   \mu - Q_k \la)   \exp \B( - \frac12\| d \cA \|^2 +<     Z , J> \B)\    D\cA\  D Z\  D \la\   D \mu \\
\end{split}
\ee   
\bigskip

\noindent  \textbf{step 3:}
Now  make the change of variables 
\be   \la'   =  \la +   \la_0   \ee
where 
$\la_0$ is chosen   so that   $Q_k \la_0  = \mu$  and   $R_k \De \la_0  =0$.
These equation have a unique solution which turns out to be   \cite{Bal85b}
\be  
 \la_0  =   G_k ^2Q_k^T (Q_k G^2_k Q_k)^{-1}( I- a Q_k G_k Q^T_k) \mu  + a G_kQ_k^T \mu 
\ee
where  $G_k = ( - \De  + a Q_k^TQ_k )^{-1}$ and  $a>0$ is arbitrary.  
Then  $\de ( \mu - Q_k \la ) =  \de (Q_k \la')$ and since     $\cQ_k \la'=0$  we    have 
 $ \pa  Q_k \la  =   \pa   Q_k \la'  - \pa  \mu  = - \pa \mu$.
Also    $R_k \De \la =  R_k  \De \la' $.     Hence we get     
\be  \label{dogged4} 
\begin{split}
  &    \const   \int        \de (  \cQ   Z  )  \de(  \tau Z)  \de (Z -   \cQ_k  \cA    -  \pa \mu  )\ 
  \de_{R_k}\B( R_k  (\de \cA + \De \la' ) \B)  \de_{R_{k+1}}\B( R_{k+1} \de \cA  \B) \\
&   \de(  Q\mu  )  \de  (  Q_k \la')  \exp \B( -\frac12 \| d \cA \|^2 +<     Z , J> \B)\    D\cA\  D Z\  D \la'  D \mu \\
\end{split}
\ee   
We now have seven delta functions and the task is   to remove all of them.  
\bigskip

\noindent  \textbf{step 4:}
The integral over  $\la'  $  is now  (Proposition \ref{lunk} again)
\be
 \int        \de  (  Q_k \la') 
  \de_{R_k}\B( R_k ( \de \cA + \De \la' )  \B) D \la '  =   \const   
  \ee
So this leaves us with        
\be  \label{dogged5} 
\begin{split}
  &     \const   \int        \de (  \cQ   Z  )   \de(  \tau Z) \de (Z -   \cQ_k  \cA   -  \pa \mu  )\ 
    \de_{R_{k+1}}\B( R_{k+1} \de \cA \B )   \de(  Q\mu  )  \\
&  \exp \B( -\frac12 \| d \cA \|^2 +<     Z , J> \B)\    D\cA\  D Z\   D \mu \\
\end{split}
\ee   
\bigskip

\noindent  \textbf{step 5:}
 Next   make the gauge transformation  $Z \to  Z +  \pa    \mu$.   This leaves    $  \de (  \cQ   Z  )  $ invariant
 since  $Q \mu =0$  and we have    
  \be  \label{dogged6} 
\begin{split}
  &     \const   \int        \de (  \cQ   Z  )   \de\B(  \tau (Z+ \pa  \mu)\B) \de (Z -   \cQ_k  \cA  )\ 
    \de_{R_{k+1}}\B( R_{k+1} \de \cA  \B)   \de(  Q\mu  )  \\
& \exp \B( -\frac12 \| d \cA \|^2 +<     Z +  \pa   \mu   , J> \B)\    D\cA\  D Z\   D \mu \\
\end{split}
\ee   
\bigskip   

\noindent  \textbf{step 6:}  In general  for $Z, \mu$ on  a unit lattice   
let      $\mu = \cM Z $ be the solution of the 
 equations  for  $x \in B(y)$
 \be    ( \tau  ( Z + \pa  \mu  ) ) (y,x)  =  0   \ \ \   x \neq y    \hs   Q \mu (y)   = 0      \ee       
 This can also be written 
 \be    ( \tau  Z  ) (y,x)   +   \mu (x) - \mu (y)     =  0   \ \ \   x \neq y    \hs   \mu (y)   =  - \sum_{x' \neq y } \mu(x')      \ee   
 The  
 solution is  
 \be  
 \mu(x)  =  \cM  Z(x)   =
- ( \tau  Z  ) (y,x)   +    L^{-3} \sum_{x' \neq y }( \tau  Z  ) (y,x')    
\ee
The  integral over $\mu$    in  (\ref{dogged6})     is        
\be
\int       \de\B(  \tau (Z+ d \mu)\B)  \de(  Q\mu  )  
  \exp \B( <     Z +  \pa   \mu   , J> \B)   D \mu \
\ee
The  delta functions     select        $\mu  =  \cM Z $    and so  (\ref{dogged6}) becomes
  \be  \label{dogged7} 
\begin{split}
  &       \const     \int        \de (  \cQ   Z  )   \de (Z -   \cQ_k  \cA  )\ 
    \de_{R_{k+1}}\B( R_{k+1} \de \cA \B )  
 \exp \B( -\frac12 \| d \cA \|^2 +<     Z +  \pa \cM Z   , J> \B)\    D\cA\  D Z\    \\
\end{split}
\ee   
Now    do the integral over  $Z$  and get 
  \be  \label{dogged8} 
\begin{split}
  &      \const     \int        \de (    \cQ_{k+1}  \cA   )   \ 
    \de_{R_{k+1}}\B( R_{k+1} \de \cA  \B)    
 \exp \B( -\frac12 \| d \cA \|^2 +<    (I+ \pa \cM )  \cQ_k  \cA    , J> \B)\    D\cA\  \\
\end{split}
\ee   
\bigskip

\noindent  \textbf{step 7:}
Next   we  change from the delta function  gauge fixing    $ \de_{R_{k+1}}( R_{k+1} \de \cA )    $ 
to   exponential    gauge fixing given by    $  \exp  ( - \frac12 \| R_{k+1} \de \cA  \| ^2 )$.
The cost is that we make the gauge  transformation 
\be
\cA  \to   \cA -   \pa  G_{k+1}R_{k+1}\de \cA 
\ee 
  ($R_{k+1}, G_{k+1}$ are  still on  $\tk$.)  This is a Fadeev-Popov argument;  see Proposition \ref{pop} in      the appendix for the details.  
But this  particular     gauge transformation changes nothing in  (\ref{dogged8})  as  we  now explain.

First we claim  that   $Q_kG_k R_k  =0$.    Indeed    for  any  scalar  $f$,  $R_k f  =  \De  \la$ for some  $\la$ satisfying  $Q_k \la =0$
and  then  
\be   Q_k G_k R_k f  =   Q_k G_k  \De \la   =   -  Q_k G_k ( - \De + a  Q_k^T Q_k)  \la = - Q_k \la  =0
\ee
Hence the change in     $\cQ_{k+1} \cA$   under the gauge transformation is   
\be  \cQ_{k+1}    \pa G_{k+1}R_{k+1}\de \cA 
=  \pa    Q_{k+1} G_{k+1}R_{k+1}\de \cA  =0
\ee
The change in     $( I + \pa  \cM )  \cQ_k  \cA $  under the gauge transformation is
\be   
( I + \pa  \cM )  \cQ_k     \pa  G_{k+1}R_{k+1}\de \cA   =   ( I + \pa   \cM )\pa   Q_k G_{k+1}R_{k+1}\de \cA     =0   
\ee
since $Q  (Q_k G_{k+1}R_{k+1}\de \cA )    =   Q_{k+1} G_{k+1}R_{k+1}\de \cA   =0$
and in general if  $Q  \mu  =0  $  then  $ \cM  \pa   \mu   = -   \mu $.

Thus our expression has become   
 \be  \label{dogged9}  
   \const     \int        \de (    \cQ_{k+1}  \cA   )   \ 
 \exp \B( -\frac12 \| d \cA \|^2  - \frac12 \| R_{k+1} \de \cA  \| ^2+<   (I+ d \cM )  \cQ_k  \cA,     J> \B)\    D\cA\ 
 \ee   
\bigskip

\noindent  \textbf{step 8:}
The last integral has the form         
\be    \label{swamped}  
   \int        \de (    \cQ_{k+1}  \cA   )   \ 
 \exp \B( -\frac 12  \| d \cA \|^2  - \frac12 \| R_{k+1} \de \cA  \| ^2+<  \cA,  J > \B)\    D\cA\  
\ee  
Note that  for any  $a>0$    we     can insert   a term   $-\frac12  \| \cQ_{k+1} \cA  \|^2  $  in the exponential.
By computing the minimizer  in  $\cA$ of   the exponential subject to $\cQ_{k+1} \cA   =0  $
one finds that   it  is  $\cA  =  \tilde \cG_{k,0} J$  where   
\be    \label{lorenzo}    
\tilde \cG_{k,0}  =   \cG_{k,0} -  \cG_{k,0}\cQ_{k+1}^T   \B(\cQ_{k+1} \cG_{k,0}\cQ_{k+1}^T\B)^{-1}   \cQ_{k+1} \cG_{k,0}
\ee
and   where     $\cG_{k,0}$   (also     called  $\cG^0_{k+1}$ since it scales to $\cG_{k+1} $)     is the operator on functions on  bonds in  $\tk$:     
\be  
\cG_{k,0}   = \B(   \de d   +    d R_{k+1} \de   + a \cQ^T_{k+1} \cQ_{k+1}  \B)^{-1}
\ee
Here in $\de d$ the operator  $\de = d^T$ acts on functions on plaquettes.
We  can also  write     
\be 
   \tilde \cG_{k,0}   =  \cG_{k,0}^{\frac12}   \B[ I -  \cG_{k,0}^{\frac12} \cQ_{k+1}^T   \B(\cQ_{k+1} \cG_{k,0}\cQ_{k+1}^T\B)^{-1}   \cQ_{k+1} \cG_{k,0}^{\frac12}  \B]     \cG_{k,0}^{\frac12}  
\ee
The   bracketed  expression is  identified as a projection operator,   and hence  
\be      \label{slow} 
    \tilde \cG_{k,0}  \cG_{k,0}^{-1}    \tilde \cG_{k,0}  =     \tilde \cG_{k,0} 
 \ee 
 The integral   (\ref{swamped})  can be  written   $   \int        \de (    \cQ_{k+1}  \cA   )   \ 
 \exp \B( - \frac12  <   \cA,    \cG_{k,0}^{-1}  \cA > +<  \cA,  J > \B)\    D\cA\  $.  Expanding  around the minimum 
 $\cA  =  \tilde \cG_{k,0} J$    and using the identity  (\ref{slow})   we find 
 \be 
   (\ref{swamped})    =   \const     \exp \B(  \frac12 < J, \tilde   \cG_{k,0}  J>   \B)    
 \ee
Therefore        
  (\ref{dogged9})  can be written    
 \be  \label{dogged10} 
    \const   \      \exp \B(    \frac12   <  \cQ^T_k (I  +    \pa  \cM)^TJ,\   \tilde      \cG_{k,0}      \cQ^T_k(I  +    \pa  \cM)^T   J> \B) 
\ee   
This gives the  desired representation:
 \begin{prop}
   \be     CC_k C^T   =   (I +  \pa \cM) \cQ_k \tilde      \cG_{k,0}       \cQ^T_k(I  +   \pa \cM)^T \ee
\end{prop}  
 \bigskip

 \subsection{representation for $   C_k^{\frac12}  $ }
 
 It will  also     be      useful to have   a representation for      $C_k^{\frac12} $  or   $C  C_k^{\frac12}  C^T$  following   \cite{Bal88a}.  
 Start with  $ C_k    =   ( C^T \De_k  C  )^{-1}$  and     represent the square root
 as        
 \be  
 \label{picnic}
         C_k^{\frac12}  =      \frac{1}{\pi}  \int_0^{\infty}  \frac{dx}{ \sqrt x  }      C_{k,x}  
\hs   
   C_{k,x}  =    \B( C^T    \De_k C   +x\B)^{-1}  
\ee
It  is sufficient then to find a representation for   $ C_{k,x}  $ 
or  $ C   C_{k,x}  C^T $.

Following the proof  for    $ C   C_{k}  C^T $  
we start with         
\be
\begin{split}
 \exp \B( \frac 12  <J, C     C_{k,x}   C^T  J>  \B )    =   \const   &   \int    e^{   <C \tilde   Z , J>  }
   \exp \B(  - \frac12   <C  \tilde   Z , \De_k C  \tilde   Z >  - \frac{x}{2}   \|  \tilde   Z \|^2     \B)    \    D \tilde   Z   \\
\end{split}
\ee   
But  $C \tilde   Z  = ( \tilde   Z,  S( \tilde Z) )$ where  $S \tilde Z $   is  defined on the central linking  bonds  $b(y,y')$
Thus if  $\chi^*$ is the characteristic function of  $\bbT^0_{N-k}  -  \{  b(y,y')  \}  $ we have
$ \tilde Z  =  \chi^*  C \tilde Z  $.     Then we can write     the integral as    
\be   
 \const   \int   e^{  <     Z , J>  } 
      \de (  \cQ   Z  )   \de(  \tau Z)  \     \exp \B(   - \frac12 <Z,  \De_k Z >  - \frac{x}{2}   \|\chi^* Z \|^2     \B)  D Z
\ee   
 The only difference  from the previous lemma     is  the term    $ - \frac{x}{2}   \|\chi^* Z \|^2   $ in the exponential.  
 This has no effect  up to   step 4   which now reads   
   \be  \label{kitty} 
\begin{split}
  &     \const   \int        \de (  \cQ   Z  )   \de(  \tau  Z) \de (Z -   \cQ_k  \cA - \pa \mu )\ 
    \de_{R_{k+1}}\B( R_{k+1} \de \cA  \B)   \de(  Q\mu  )  \\
& \exp \B( - \frac12 \| d \cA \|^2  - \frac{x}{2}   \|\chi^* Z \|^2  +<     Z  , J> \B)\    D\cA\  D Z\   D \mu \\
\end{split}
\ee   
In subsequent steps  we  have     $Z \to Z  +   \pa     \mu$ ,  then    $\mu  =   \cM   Z  $,  and then    $Z=    \cQ_{k}  \cA $.
This   brings us to    
   \be  \label{kitty5} 
   \begin{split}
  &    \const     \int     \    D\cA\     \de (    \cQ_{k+1}  \cA   )   \\
  & 
 \exp \B( -  \frac12 \| d \cA \|^2  - \frac12 \| R_{k+1} \de \cA  \| ^2-  \frac{x}{2}  \|\chi^*  (I + \pa \cM )   \cQ_k  \cA   \|^2
 +< (I + \pa    \cM )   \cQ_k  \cA,   J> \B)   \\
 \end{split}
\ee   
Define   $\tilde \cG_{k,x} $  by   
\be 
\begin{split}
 &  \exp \B( \frac12 <f,  \tilde \cG_{k,x}  J>  \B )\\  =   
 &   \const     \int        \de (    \cQ_{k+1}  \cA   )   \ 
 \exp \B( -  \frac12 \| d \cA \|^2  - \frac12 \| R_{k+1} \de \cA  \| ^2-  \frac{x}{2}  \| \chi^* (I + \pa    \cM )   \cQ_k  \cA   \|^2+<  \cA,   J> \B)\    D\cA\\
 \end{split} 
 \ee
 The minimizer   of the exponential  subject to  $  \cQ_{k+1}  \cA=0  $     is   $\cA  =   \tilde       \cG_{k,x} J$ 
 where     
\be    \label{swan}
  \tilde       \cG_{k,x}   =  
    \cG_{k,x} -  \cG_{k,x}\cQ_{k+1}^T   \B(\cQ_{k+1} \cG_{k,x}\cQ_{k+1}^T\B)^{-1}   \cQ_{k+1}  \cG_{k,x}
\ee
 and     where  for any  $a>0$ 
\be   
\cG_{k,x}  = \B(   \de   d   +    d R_{k+1} \de   + a \cQ^T_{k+1} \cQ_{k+1}    +  x\cQ_k^T   (I + d \cM )^T\chi^* (I + d \cM )    \cQ_k   \B)^{-1}
\ee
Expanding around the minimizer    (\ref{kitty5}) becomes   
 \be  \label{kitty6} 
  \const     \exp \B( \frac12 < \cQ_k^T  (I  +  \pa   \cM)^T J,  \tilde \cG_{k,x} \cQ_k^T (I  +  \pa \cM)^T  J>   \B)  
\ee   
Thus we have established the representation:   
\begin{prop} 
\be     \label{sally}  
   C     C_{k,x}   C^T  =     (I  +  \pa  \cM) \cQ_k\tilde \cG_{k,x} \cQ_k^T  (I  +   \pa   \cM)^T  
    \ee
 \end{prop}
 \bigskip

\subsection{summary} 

Thanks to the parametrization of section  \ref{parametrization}   
the fluctuation integral  (\ref{owen}) can      be written as       the Gaussian  integral        
\be    \label{sometimes2} 
  F_{k+1}  (\cA)     =    \int           F_k \B( \cA_L +  \cH^{\sx}_k C    \tilde Z      \B)    d \mu_{C_k}   (  \tilde Z)      
\ee      
We  want to evaluate this at  $\cA  =  \cH^{\sx}_{k+1}  A_{k+1}$.   But  since  
  $F_k$ is gauge invariant   and    since  $\cH^{\sx}_{k}$  and  $\cH_{k}$    are  related by a gauge transformation 
we can equally we  evaluate it  at  the more regular      $\cA  =  \cH_{k+1}  A_{k+1}$   and with         $\cH^{\sx}_k C    \tilde Z  $
replaced by  $\cH_k C    \tilde Z$.    Furthermore  we  can   remove the non-locality  from the Gaussian measure  
by    the change of variables  $\tilde Z  =  C_k^{\frac12}  \tilde W $  where  $\tilde W$ is a variable of the same type as  $\tilde Z$.
  Then   we have    
\be    \label{sometimes3} 
  F_{k+1}  (\cA)     =    \int           F_k \B( \cA_L +  \cH_k C C_k^{\frac12}     \tilde W      \B)    d \mu_{I }   (  \tilde W)      
\ee     
 In  this  last form  the fluctuation  integral  is subject  to rigorous  analysis.  The  fields  $\cA  =  \cH_{k+1}  A_{k+1}$ and
$  \cH_kC  C_k^{\frac12}   \tilde Z $   have 
 good regularity properties.   The operator   $C_k^{\frac12}$  has a random walk expansion  which can be 
 derived from the representation  (\ref{sally}).   Hence it can be broken into local pieces  and
 the    integral  can then be  treated by the standard     technique  of   a cluster expansion.   
  This  means  that  if
 $  F_k$ has an expansion into local pieces,  then one  can  expand  $ F_{k+1}$ into  local  pieces.
 This  is the key  issue  in studying the mapping  $F_k  \to F_{k+1}$ and controlling the flow.   Variations of this program are carried out in     \cite{Bal88a},   \cite{BIJ88},  \cite{Dim14}.

 \appendix
 
 \section{change of gauge}
 
 We   explain how to change between    the  generalized    Landau gauge to a  generalized Feynman gauge.   First a preliminary result:  
 
 \begin{prop}   \label{lunk}   For  $\la$  on  $\tk$   and $R_k$ the projection onto  $\De  (  \ker  Q_k  ) $ and $\De = - \de  d$
  \be   \label{soused1} 
1 =    \const    \int      \de ( Q_k \la)  \exp \B( - \frac12  \|R_k \de  \cA  - \De  \la    \|^2 \B) D \la 
\ee
\be
\label{soused2}     
1 =    \const    \int      \de ( Q_k \la)  \de_{R_k} \B(R_k \de  \cA  - \De  \la  \B)    D \la 
 \ee
\end{prop}
\bigskip

\rem   Since  $Q_k \la =0$ we can replace $\De \la$  by   $R_k \De \la$ in these formulas.  
\bigskip 

\pr     It suffices to show that  the mapping    $\la  \to  (Q_k \la,   R_k \De \la)$   from    
$ \bbR^{  \tk}$   to       $ \bbR^{\bbT^0_{N-k}}  \
 \times \ran \  R_k $  
  is a bijection.   Then either     result follows by  making this change of variables.   We use  that   
  $\De  $ is a bijection from  $\ker\  Q_k $  to   $\ran\    R_k$.   
  The  mapping is injective   since
 if   $Q_k \la  =0$  and  $R_k \De \la  = 0$ then  $\De \la  =0$  and hence  $\la =0$.
 The dimensions match  since     
 \be
    \dim   (  \ran  \  Q_k  )  +  \dim  ( \ker \ Q_k  )   =   \dim  (   \bbR^{  \tk}   )
 \ee
 says that
 \be    
   \dim   (    \bbR^{\bbT^0_{N-k}}  )  +  \dim  (\ran \  R_k   )   =   \dim  (   \bbR^{  \tk}   )
\ee  
 Hence the result.

 \begin{prop}  \label{pop}  For  $G_k =  (-\De  + a Q_k^T Q_k)^{-1} $ and   any  $a \geq   0$
 \be   \label{swish}
 \begin{split}
& \int   \de (\cQ_k \cA )  \de_{R_k} \B( R_k \de \cA  \B)  f(\cA )  \exp \B( -\frac12 \| d\cA  \|^2 \B)  D \cA \\
   &   =    \const   \int      \de (\cQ_k \cA )   f\B(\cA  -  \pa   G_kR_k \de \cA \B)  \exp \B( - \frac12    \| d\cA  \|^2  - \frac12  \|R_k \de \cA \|^2 \B)  D \cA\\
\end{split}
 \ee
 \end{prop} 
 
 \pr
 Insert  (\ref{soused1}) under the integral sign on the left side  of (\ref{swish}).   
  Change the order of integration and  make the gauge transformation $\cA \to \cA^{\la}  = \cA -  \pa  \la$. 
   Then  $\de \cA  \to   \de \cA  + \De \la$ and  $\cQ_k \cA$ is invariant since   $Q_k \la = 0$.  
 This yields   
  \be
 \begin{split}
& \const    \int    \de ( Q_k \la)     
 \B[   \int   \de (\cQ_k \cA )  \de_{R_k} \B( R_k  \de \cA  + \De \la  \B)  f(\cA-   \pa   \la )   \exp \B( - \frac12    \| d\cA  \|^2  - \frac12  \|R_k \de \cA \|^2 \B)   D \cA\B]
D \la     \\
\end{split}
 \ee
 Change the order of integration again and do the integral over  $\la$.   The  delta functions in    $\la$ 
 select    
 \be       Q_k \la  =0   \hs    \De  \la   =   -   R_k  \de \cA    \ee
 or  equivalently  
 \be       Q_k \la  =0   \hs     ( - \De + a Q^T_k Q_k ) \la   = R_k  \de \cA    \ee
 The unique solution is    (recall  $Q_kG_kR_k =0$)
 \be 
   \la   =   G_kR_k \de \cA 
    \ee
    Make this replacement in  $f(\cA-   \pa   \la )$.
Then the integral over  $\la$  is constant    by  (\ref{soused2})   and  we are left with  the right side  of   (\ref{swish}).

\end{document}